\documentclass[final,5p,times,twocolumn]{elsarticle}

\usepackage{amssymb, amsmath, amsfonts}
\usepackage{xcolor}
\usepackage[list=true, position=top]{subcaption}
\usepackage{tabularx}
\usepackage{multirow}
\usepackage{booktabs}
\usepackage[capitalise]{cleveref}
\usepackage{algorithmic}
\usepackage{algorithm}
\usepackage{array}
\usepackage{textcomp}
\usepackage{stfloats}
\usepackage{url}
\usepackage{verbatim}
\usepackage{graphicx}
\usepackage{import}
\DeclareMathOperator*{\argmin}{arg\,min}

\usepackage{tikz}
\usetikzlibrary{arrows.meta, shapes.geometric, arrows}
\tikzstyle{arrow} = [arrows = {-Stealth[scale=1.2]}]

\tikzstyle{text_box} = [rectangle, minimum width=1cm, minimum height=0.5cm, text centered, draw=black]
\tikzstyle{rounded_text_box} = [rectangle, minimum width=2cm, minimum height=0.5cm, rounded corners, text centered, draw=black]
\tikzstyle{large_rounded_text_box} = [rectangle, minimum width=2.5cm, minimum height=1.5cm, rounded corners, text centered, draw=black]
\tikzstyle{circ} = [circle, minimum size=0.3cm, text centered, draw=black]
\tikzstyle{bus} = [rectangle, minimum width=2cm,minimum height=0.15cm, draw=black, rounded corners]
\tikzstyle{block} = [rectangle, minimum width=1cm, minimum height=1.5cm, text centered, draw=black]

\usepackage{pgfplotstable}
\usepackage{pgfplots}
\pgfplotsset{compat=1.12}

\usepackage{acronym}
\acrodef{pinn}[PINN]{Physics-informed Neural Network}
\acrodef{bpinn}[BPINN]{Bayesian Physics-informed Neural Network}
\acrodef{sindy}[SINDy]{Sparse Identification of Nonlinear Dynamics}
\acrodef{smib}[SMIB]{single machine infinite bus}
\acrodef{mape}[MAPE]{mean absolute percentage error}
\acrodef{nn}[NN]{Neural Network}
\acrodef{bnn}[BNN]{Bayesian Neural Network}
\acrodef{ml}[ML]{Machine Learning}
\acrodef{ibr}[IBR]{Inverter-based Resource}

\journal{Electric Power Systems Research}

\begin{document}

\begin{frontmatter}

\title{Bayesian Physics-informed Neural Networks for System Identification of Inverter-dominated Power Systems}

\author{Simon Stock$^a$, Davood Babazadeh$^a$, Christian Becker$^a$}

\affiliation{organization={Institute of Power and Energy Technology, Hamburg University of Technology},
            city={Hamburg},
            country={Germany}}

\author{Spyros Chatzivasileiadis$^b$}
\affiliation{organization={Department of Wind and Energy Systems, Technical University of Denmark},
            city={Kgs. Lyngby},
            country={Denmark}}

\begin{abstract}
While the uncertainty in generation and demand increases, accurately estimating the dynamic characteristics of power systems becomes crucial for employing the appropriate control actions to maintain their stability. In our previous work, we have shown that \acp{bpinn} outperform conventional system identification methods in identifying the power system dynamic behavior under measurement noise. This paper takes the next natural step and addresses the more significant challenge, exploring how \ac{bpinn} perform in estimating power system dynamics under increasing uncertainty from many \acfp{ibr} connected to the grid. 
These introduce a different type of uncertainty, compared to noisy measurements. 
The \ac{bpinn} combines the advantages of \acp{pinn}, such as inverse problem applicability, with Bayesian approaches for uncertainty quantification. 
We explore the \ac{bpinn} performance on a wide range of systems, starting from a \ac{smib} system and 3-bus system to extract important insights, to the 14-bus CIGRE distribution grid, and the large IEEE 118-bus system. We also investigate approaches that can accelerate the \ac{bpinn} training, such as pretraining and transfer learning. Throughout this paper, we show that in presence of uncertainty, the \ac{bpinn} achieves orders of magnitude lower errors than the widely popular method for system identification \acs{sindy} and significantly lower errors than \ac{pinn}, while transfer learning helps reduce training time by up to $80\,\%$. 
\end{abstract}




\end{frontmatter}

\section{Introduction}
\label{sec:introduction}
The ongoing integration of \acfp{ibr} into the energy system leads to significant changes in frequency dynamics, since they do not show the same characteristics as conventional synchronous generators. Additionally, the volatile behavior of wind power plants and PV causes a fluctuating infeed of power that increases uncertainty in grid operation. 
For that reason, system operators appear to lack system awareness in case of high penetration of renewables. Nevertheless, the introduction of fast and distributed measurement devices, namely phasor-measurement-units (PMUs), enables the utilization of data-driven approaches for identification of system dynamics. 

Various approaches have been introduced, such as filter-based techniques, e.g. Kalman filtering \cite{zhao2019}, Koopman theory \cite{Susuki2018} or parsimonious approaches, for example \ac{sindy} \cite{Brunton2016}. 
Many of them have shown vulnerability to uncertainty in the data, resulting in inaccurate estimates. Machine learning and, more recently, hybrid approaches that combine the strengths of machine learning with physics-based models have been introduced to solve this problem, for example the \acfp{pinn} \cite{Stiasny2020}. However, \acp{nn} and \acp{pinn} do not inherently quantify the experienced uncertainty and therefore lack a confidence measure about their estimate. In Bayesian techniques, the estimate is augmented with such a confidence measure by design \cite{Petra2017}.   

In that sense, the \acf{bpinn} has been introduced which combines the \ac{pinn} and Bayesian techniques \cite{YANG2021}. In power systems, it has shown robustness against uncertainty from noisy measurements for system identification \cite{Stock2023}, outperforming established approaches, such as \ac{sindy}. This type of uncertainty is commonly known as aleatoric uncertainty. In this paper, we further evaluate the \ac{bpinn} as follows: First, we investigate the \ac{bpinn} performance for estimating the frequency dynamics of an inverter-dominated grid. \acp{ibr} lead to model uncertainties in system identification, which are commonly known as epistemic uncertainty. Second, we seek to transfer previously learned knowledge in \ac{bpinn} training. After pretraining on a \ac{smib} system, we transfer the knowledge and train on a larger system in order to reduce the required data and training iterations.

Most Bayesian approaches require informative prior knowledge about the inferred system parameters \cite{Petra2017}. These are expected to change frequently for inverter-dominated power systems, thus, we focus on weakly-informative priors. In contrast to informative priors, weakly-informative priors are generally applicable to the whole range of system parameters. 

The rest of this paper is structured as follows: We start by introducing the methodology of the \ac{bpinn} in \cref{sec:methodology}, highlight the similarities to the \ac{pinn}, and detail its uncertainty quantification capabilities. We then specify the weakly-informative priors and simulation parameters. In \cref{sec:modeling/case_study}, the \ac{ibr} and generator models are presented and the four grid models are specified. \cref{sec:results} discusses the results of the system identification and transfer learning and compares it to the \ac{pinn} and \ac{sindy}. \cref{sec:conclusion} concludes this paper.

\section{Methodology}
\label{sec:methodology}
We start this section by revisiting the \ac{nn} and extend it to the \ac{pinn} formulation. Based on that, we present the \ac{bpinn} and its uncertainty quantification capabilities.

Let us assume a dynamic model described by the following set of differential equations: 
\begin{align}
    \Dot{\boldsymbol{x}}&=f(\boldsymbol{x}, \boldsymbol{u};\boldsymbol{\lambda})
    \label{eq:dae_system}
\end{align}
with solution $\boldsymbol{x}(t,\boldsymbol{u})$, $\boldsymbol{x}$ representing the states and $\boldsymbol{u}$ the inputs of the system. $\boldsymbol{\lambda}$ describes the system parameters, e.g. the damping constant of a mass-spring oscillator, and operator $f$ maps the system parameters to the states.

\subsection{Neural Networks}
\acp{nn} can generally be used as function approximators for a variety of problems. For dynamic systems, they can serve as a surrogate model $g(t)$ mapping a time-dependent input vector to the target trajectory of the states $\boldsymbol{x}$:
\begin{align}
    g(t; \boldsymbol{\Theta}) =\hat{\boldsymbol{x}}(t; \boldsymbol{\Theta})\approx \boldsymbol{x}(t, \boldsymbol{u}; \boldsymbol{x_0}, \boldsymbol{\lambda}).
    \label{eq:surrogate_model}
\end{align}
$\boldsymbol{x_0}$ describes the initial state of the system and $\boldsymbol{\Theta}$ the \acl{nn} parameters, i.e. the \ac{nn} weights and biases.
A set of measurement data $\mathcal{D}$ with $\mathcal{D}=\{\boldsymbol{x}^{(i)}, \boldsymbol{u}^{(i)} \}_{i=1}^{N_z}$ can be used to determine the surrogate model parameters $\boldsymbol{\Theta}$ through training the \ac{nn}. 
The distance between the estimate $\hat{\boldsymbol{x}}(t)$ and the target trajectory is minimized by updating the \ac{nn} parameters $\boldsymbol{\Theta}$. \cref{eq:forward_optimization_NN} formulates this as an optimization problem using the root-mean-squared error as the distance measure. 
This \acf{nn} training procedure is a supervised learning problem, i.e. it requires a fully labeled dataset $\mathcal{D}$ including all true values $\boldsymbol{x}$.
\begin{align}
    \min_{\Theta}  \frac{1}{N_z} \sum_{i=1}^{N_z}\sqrt{(\hat{\boldsymbol{x}}^{(i)}(\Theta)-\boldsymbol{x}^{(i)})^2}
    \label{eq:forward_optimization_NN}
\end{align}
This formulation strives to find the optimal model parameters $\boldsymbol{\Theta}$. However, it is not able to obtain the system parameters $\boldsymbol{\lambda}$. An inverse problem has to be formulated to determine the system parameters $\boldsymbol{\lambda}$ as well. To this end, the \acf{pinn} has been proposed \cite{RAISSI2019}. \acp{pinn} incorporate a physics regularization term into the loss function, as follows:

\begin{align}
    h(t,\boldsymbol{u}; \boldsymbol{\Theta}, \boldsymbol{\lambda})&=\frac{d}{dt}\hat{\boldsymbol{x}}-f(\hat{\boldsymbol{x}}, \boldsymbol{u};\hat{\boldsymbol{\lambda}}) \stackrel{!}{=} 0.
\label{eq:residual_loss}
\end{align} 
This formulation is based on the differential equation that describes the dynamic model \cref{eq:dae_system} and augments \cref{eq:forward_optimization_NN} in the form $\frac{1}{N}\sqrt{h^2}$. $\boldsymbol{\Theta}$ can now be tuned to find the state estimates $\boldsymbol{\hat{x}}$ and system parameter estimate $\boldsymbol{\lambda}$ collaboratively. 
This formulation does not require labeled data for $\boldsymbol{\lambda}$, while it still necessitates labels for $\boldsymbol{x}$. 
 
The evaluation of \cref{eq:residual_loss} can be extended with additional data that are not in the measurement data $\mathcal{D}$. The \ac{nn} \cref{eq:surrogate_model} can be evaluated at every point in time $t$ to generate $N_c$ so-called collocation points so that the total number of training points is $N=N_z+N_c$. The $N_c$ collocation points can support the physics regularization in \cref{eq:residual_loss}. In conclusion, the \ac{pinn} enables estimating the measurement trajectory $\boldsymbol{x}$ and system parameters $\boldsymbol{\lambda}$ at the same time based on given measurement data $\mathcal{D}$, however, it does not indicate its estimates uncertainty.

\subsection{Bayesian PINN}
\label{subsec:bpinn}
Bayesian frameworks have been introduced to deep learning techniques to quantify the uncertainty in measurements and modeling \cite{Kendall2017}. The literature mostly distinguishes between two types of uncertainty: first, aleatoric uncertainty, which represents noise inherent in the observed data and second, epistemic uncertainty, which describes uncertainty in the model. 
\paragraph{Uncertainty quantification}
Aleatoric uncertainty is commonly assumed to be Gaussian distributed. An artificial set of observed noisy data $\mathcal{D}$ can be created with a deterministic process, such as \cref{eq:dae_system}, giving the mean and additive noise provided by a covariance matrix $\Sigma_x=\sigma^2_x\boldsymbol{I}$. This type of uncertainty can potentially be quantified by extending \cref{eq:forward_optimization_NN} as follows \cite{Kendall2017}: 
\begin{align}
    \min_{\boldsymbol{\Theta}} \frac{1}{N} \sum_{i=1}^{N}\frac{1}{2\sigma_x^{(i)^2}}\sqrt{(\hat{\boldsymbol{x}}^{(i)}(\boldsymbol{\Theta})-\boldsymbol{x}_{\text{true}}^{(i)})^2}+\frac{1}{2}\text{log}\,\sigma_x^{(i)^2}.
    \label{eq:aleatoric_loss}
\end{align}
Note that this formulation does not lead to a Bayesian \ac{nn}, which means that we have to find a single value for each neural network parameter $\Theta$.

\acp{bnn} were introduced to address the problem of epistemic uncertainty, resulting from the model, about three decades ago \cite{Kononenko1989}. A probability distribution $p(\boldsymbol{\Theta})$ is placed on the model parameters $\boldsymbol{\Theta}$ based on prior beliefs. In training, their posterior distributions $p(\boldsymbol{\Theta}|\mathcal{D})$ are inferred using Bayes' rule and the data $\mathcal{D}$. Subsequently, the \acf{bnn} can be envisioned as a family of models that incorporates the plausible set of parameters.
The posterior distribution of model parameters $\boldsymbol{\Theta}$ can be obtained based on 
\begin{align}
    p(\boldsymbol{\Theta|}\mathcal{D})=\frac{p(\mathcal{D}\boldsymbol{|\Theta}) p(\boldsymbol{\Theta})}{p(\mathcal{D})}
\end{align}
using a Bayesian inference algorithm. 
Samples can be pulled from the inferred parameter distribution $\boldsymbol{\Theta}^* \sim p(\boldsymbol{\Theta}|\mathcal{D})$. These can be used to calculate the mean of the distribution of estimated states by 
\begin{align}
\mathbb{E}[\hat{\boldsymbol{x}}|\boldsymbol{u},\mathcal{D}]\approx \frac{1}{T} \sum_i^T \hat{\boldsymbol{x}}_{\Theta_i^*}(u):=\overline{\boldsymbol{x}}_{\mathcal{D}}(\boldsymbol{u})    
\end{align} 
and the aleatoric uncertainty as $\mathbb{E}_{\Theta|\mathcal{D}}[\text{Var}(\hat{\boldsymbol{x}}|\boldsymbol{u},\boldsymbol{\Theta})]$. The aleatoric uncertainty consists of the expected variance of $\boldsymbol{\hat{x}}$, in contrast, the epistemic uncertainty can be formulated as the variance of the expected value of $\boldsymbol{\hat{x}}$. This gives us the total uncertainty \cite{graf2022}:
\begin{align}
\begin{split}
    \text{Var}(\hat{\boldsymbol{x}}|\boldsymbol{u},\mathcal{D})&=\mathbb{E}_{\boldsymbol{\Theta}|\mathcal{D}}[\text{Var}(\hat{\boldsymbol{x}}|\boldsymbol{u},\boldsymbol{\Theta})]+\text{Var}_{\boldsymbol{\Theta}|\mathcal{D}}(\mathbb{E}[\hat{\boldsymbol{x}}|\boldsymbol{u},\boldsymbol{\Theta}]).
\end{split}
\label{eq:uncertainty_quant}
\end{align}
This formulation assumes that the distribution $p(\hat{\boldsymbol{x}}|\mathcal{D})$ follows the same distribution as the observed data $\mathcal{D}$.
Although we can disassemble \cref{eq:uncertainty_quant} into its aleatoric and epistemic elements in the form of equations, the \ac{bnn} cannot distinguish between the sources of uncertainty by design, so it only provides one uncertainty measure.

The above formulations enable the \ac{bnn} to discover the state estimates $\boldsymbol{\hat{x}}$ considering the aleatoric and epistemic uncertainty based on:
\begin{align}
    p(\mathcal{D}|\boldsymbol{\Theta})=\prod_i^N\frac{1}{\sqrt{2 \pi \sigma_x^{(i)^2}}}\text{exp}\left(-\frac{(\hat{\boldsymbol{x}}^{(i)}(\boldsymbol{\Theta})-\boldsymbol{x}_{\text{true}}^{(i)})^2}{2 \sigma_x^{(i)^2}}\right).
     \label{eq:BNN_loss}
\end{align}
Please note, that we do not assume the fidelity of the measurement data to be known a priori, that would require additional information to the measurement data. $\sigma_x$ has to be determined during training. 

\begin{figure}[!ht]
    \centering
        
    \begin{tikzpicture}[node distance=1.4cm]
        \node (Inp1) [circ]{$\boldsymbol{u},t$};

        \node (n21) [circ, right of=Inp1]{$\eta$};
        \node (n11) [circ,  above of=n21]{$\eta$};
        \node (nn1) [circ,  below of=n21]{$\eta$};

        \node (n12) [circ, right of=n11]{$\eta$};
        \node (n22) [circ, right of=n21]{$\eta$};
        \node (nn2) [circ, right of=nn1]{$\eta$};

        \node (Out) [circ, right of=n22]{$\hat{\boldsymbol{x}}$};
        \coordinate[right of=Out](misc4);
        \node (ddt)[circ, above of=misc4]{$\frac{\partial}{\partial t}$};
        \node (ddt2)[circ, below of=misc4]{$I$};
        \node (phys_loss) [text_box, right of=misc4]{$h(\boldsymbol{\Theta},\hat{\boldsymbol{\lambda}})$};
        \node(loss)[text_box, below of=ddt2]{$p_{\text{total}}=p_{\text{Data}}p_{\text{Physics}}$};
        \node(estimates)[text_box, left of=loss, align=left, xshift=-1.5cm]{estimates of $\boldsymbol{\lambda}$};
        \draw[dashed](loss)--(estimates);

        \coordinate [below of=Out, yshift=-1.0cm] (misc1);
        \coordinate [below of=phys_loss, yshift=-1.0cm] (misc2);
        \coordinate[above of=loss, yshift=-1cm](misc3);
        
        \draw[arrow](Inp1)--(n11);
        \draw[arrow](Inp1)--(n21);
        \draw[arrow](Inp1)--(nn1);
        \draw[arrow](n11)--(n12);
        \draw[arrow](n11)--(n22);
        \draw[arrow](n11)--(nn2);
        \draw[arrow](n21)--(n12);
        \draw[arrow](n21)--(n22);
        \draw[arrow](n21)--(nn2);
        \draw[arrow](nn1)--(n12);
        \draw[arrow](nn1)--(n22);
        \draw[arrow](nn1)--(nn2);
        
        \draw[arrow](n12)--(Out);
        \draw[arrow](n22)--(Out);
        \draw[arrow](nn2)--(Out);
        
        \draw[arrow](Out)--(ddt);
        \draw[arrow](Out)--(ddt2);
        \draw[arrow](ddt)--(phys_loss);
        \draw[arrow](ddt2)--(phys_loss);
        
        \draw[dashed](Out)--(misc1);
        \draw[dashed](phys_loss)--(misc2);
        \draw[dashed](misc1)--(misc3);
        \draw[dashed](misc2)--(misc3);
        \draw[dashed](misc3)--(loss);

        \node[align=left] at(2.1,2.5){BNN$(u,t;\Theta)$};
        \node[align=left] at(5.7,2.5){Physics};
        
    \end{tikzpicture}
    
    \caption{Bayesian Phyiscs-informed Neural Network schematic with nonlinear activation $\eta$}
    \label{fig:BPINN_structure}
\end{figure}
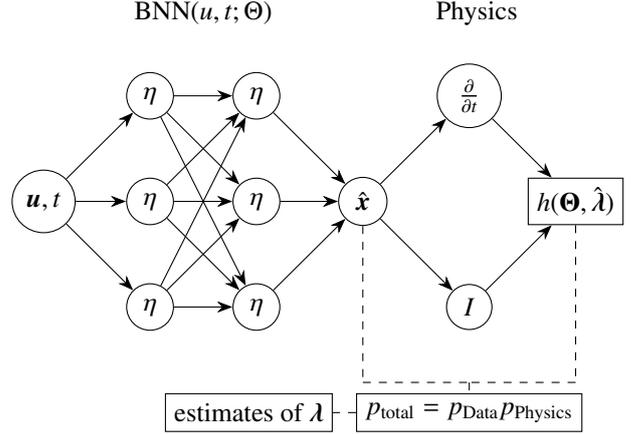

The formulation \cref{eq:BNN_loss} is solely applicable to forward problems, so we are only able to estimate the system states $\boldsymbol{x}$. Simultaneous estimation of the system parameters $\boldsymbol{\lambda}$ requires the extension of the formulation. A physical regularization similar to \cref{eq:residual_loss} is introduced, making the \ac{bnn} a \acf{bpinn} \cite{YANG2021}. This leads to the following equation: 
\begin{align}
    p_{\text{total}}(\mathcal{D}| \boldsymbol{\Theta}, \boldsymbol{\lambda})= p(\mathcal{D}|\boldsymbol{\Theta}) \prod_i^N\frac{1}{\sqrt{2 \pi \sigma_h^{(i)^2}}}\text{exp}\left(-\frac{(h^{(i)}(\boldsymbol{\Theta}, \boldsymbol{\lambda}))^2 }{2 \sigma_h^{(i)^2}}\right).
    \label{eq:BPINN_loss}
\end{align}
The \ac{bpinn} structure is illustrated in \cref{fig:BPINN_structure}.
Based on \cref{eq:BPINN_loss}, the joint posterior of the \ac{bnn} parameters $\boldsymbol{\Theta}$ and system parameters $\boldsymbol{\lambda}$ can be determined following Bayes' theorem with prior distributions $p(\boldsymbol{\Theta)}$ and $p(\boldsymbol{\lambda)}$.
\begin{align}
    p(\boldsymbol{\Theta, \lambda|}\mathcal{D})=\frac{p(\mathcal{D}\boldsymbol{|\Theta, \lambda}) p(\boldsymbol{\Theta, \lambda})}{p(\mathcal{D})}.
\label{eq:posterior_inverse}
\end{align}
We use Variational Inference (VI) to find the joint posterior of $\boldsymbol{\Theta}$ and $\boldsymbol{\lambda}$, which provides a computationally efficient formulation, that can be solved with common optimization libraries. In this paper, we specifically rely on the Stein Variational Gradient Descend (SVGD) algorithm to solve this task \cite{LIU2016}.

In conclusion, the \ac{bpinn} estimates the system states $\boldsymbol{\hat{x}}$ and system parameters $\boldsymbol{\lambda}$, while inherently indicating the uncertainty of the estimated value considering the aleatoric and epistemic uncertainty.

\paragraph{Priors for BPINNs}
The calculation of \cref{eq:BPINN_loss} requires to set a distribution $p(\boldsymbol{\Theta}, \boldsymbol{\lambda})$ based on prior beliefs. A commonly used prior $\boldsymbol{\Theta}$ for \ac{bnn} parameters is a Gaussian distribution with zero mean, standard deviation $\sigma_{w,l}=1$ and $\sigma_{b,l}=1$ for weights and biases $w_l$ and $b_l$. When the same prior is used for system parameters $\boldsymbol{\lambda}$, prior knowledge of the range of $\boldsymbol{\lambda}$ can be required. For inverter-dominated power systems, the system parameters are expected to change frequently. Hence, it is difficult to constantly update the prior beliefs.
In this paper, we use a generic prior for the system parameters $\boldsymbol{\lambda}$ that does not require informative knowledge. The so-called weakly-informative priors seek to include as little information as possible. They are based on scale mixtures of normals \cite{Lemoine2019}. Specifically, we apply the normal-gamma distribution, whose probability density function can be expressed as:
\begin{align}
    \boldsymbol{\lambda}_{prior} \sim \mathcal{N}(\mu,\kappa/\iota), \iota \sim \Gamma(\beta, \alpha).
    \label{eq:priors_for_lambda}
\end{align}

\begin{figure}
    \centering
    \input{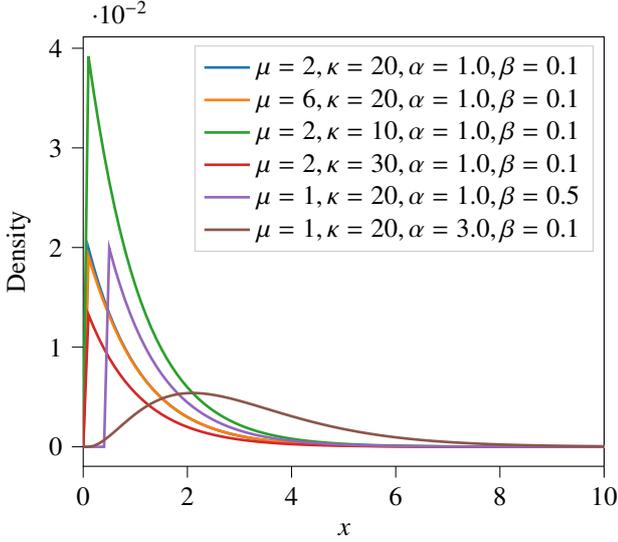}
    \caption{Exemplary normal-gamma PDFs for different parameters $\mu$, $\kappa$, $\alpha$, $\beta$}
    \label{fig:example_normal_gamma_distribution}
\end{figure}

The normal-gamma has a spike close to zero, similar to the Laplace distribution, and thus allows stronger regularization than the normal distribution. More importantly, the parameters $\alpha$ and $\beta$ can be used to control the information content of the distribution \cite{Lemoine2019}. Exemplary probability density functions for a normal-gamma distribution are shown in \cref{fig:example_normal_gamma_distribution}.
We briefly explore these parameters in \cref{sec:results} to choose the distribution parameters for this paper.

\subsection{System identification of power systems}
\label{subsec:sys_ID_eq}
In this paper, we estimate the dynamic frequency behavior of an inverter-dominated power system. This can generally be approximated by a \ac{smib} representation \cite{Bergen1981}
\begin{align}
    \dot{\delta}&=\Delta\omega\\
    \Delta\dot{\omega}&=\frac{1}{m}(P_m -d\Delta\omega-B \text{sin}(\delta)).
    \label{eq:regression_problem}
\end{align}
$m$ represents the system inertia, $d$ the damping, $B$ the susceptance and $P_m$ the mechanical power. The voltage is assumed to be one. The states of the system are the angle and the frequency deviation $\boldsymbol{x}=\{\delta, \Delta\omega\}$. The \ac{smib} representation neglects various effects when representing a more complex system with numerous \acp{ibr}. However, we will utilize it for the physics regularization in \cref{eq:BPINN_loss} throughout the entire paper, since it still properly describes the general frequency behavior. 
Consequently, \ac{bpinn} provides two distribution estimates for the system states $\boldsymbol{x}$ and three distributions for the system parameters $\boldsymbol{\lambda}=\{m,d,B\}$. These can represent a larger system, such as 118-bus, in an aggregated way.

Finally, the \ac{bpinn} parameter estimates $\hat{\boldsymbol{\lambda}}=\{\hat{m},\hat{d},\hat{B}\}$ can be obtained by taking the mean of the posterior distribution and a measure of confidence is given based on the posterior variance, as described in \cref{subsec:bpinn}. The latter indicates if the \ac{bpinn} is confident about the given estimate. 

All simulations start from an unperturbed state, so $\dot{x}=0$. We perturb the system at $t=0$ by applying a constant change in $P_m$, so $P_m=-0.1\,\text{p.u.}$.

For all tests, a \ac{bpinn} with 20 neurons, 1 hidden layer, and a standard trajectory length of $T= 5\, \text{s}$ was used at a sampling frequency of $20\,\text{Hz}$, giving $N_z=100\,\text{samples}$. The inputs of the \ac{bpinn} are $t$ and $P_m$.  
The \ac{bpinn} was implemented in Python using packages $pytorch$ and $numpyro$.

\subsection{SINDy algorithm}
\label{sec:SINDy}
The \ac{sindy} algorithm, proposed in \cite{Brunton2016}, is part of the recently popular parsimonious approaches. These focus on the active terms in a given set of differential equations in order to reduce the computational effort. A library of candidate functions is defined $\boldsymbol{\zeta}(\boldsymbol{x})$, which can be polynomial combinations of the system states $\boldsymbol{x}$. These are used to formulate a set of differential equations that represent the behavior of the system. \ac{sindy} now strives to reduce the number of equations and identify a sparse system representation $\boldsymbol{\zeta}(\boldsymbol{x})\boldsymbol{\Xi}$, with $\boldsymbol{\Xi}$ being the coefficients. Linear regression is utilized and the number of active equations is penalized through an additional term:
\begin{align}
    \argmin_{\Xi}||\dot{\boldsymbol{x}}-\boldsymbol{\zeta}(\boldsymbol{x})\boldsymbol{\Xi}||_2+\nu \boldsymbol{||\Xi||}_1. 
\end{align}
$\dot{\boldsymbol{x}}$ represents the derivatives of the states similar to previous formulations. $\boldsymbol{\zeta}(\boldsymbol{x})\boldsymbol{\Xi}$ is the dynamic system with candidate functions $\boldsymbol{\zeta}(\boldsymbol{x})$. In the estimation phase, \ac{sindy} aims to find the coefficient vector $\boldsymbol{\Xi}$ that minimizes the first term of the equation. These coefficients are additionally used in a regularizing term $\nu\boldsymbol{||\Xi||}_1$ to allow for a sparse model. 
In the previously formulated estimation problem \cref{eq:regression_problem} the candidate functions are known, since we follow the \ac{smib} representation. For that reason, the number of candidate functions is fixed in \ac{sindy}, which allows us to neglect the penalizing term $\boldsymbol{\zeta}$. 
For all tests the \textit{PySINDy} library was used \cite{Kaptanoglu2022}.

\section{Case study}
\label{sec:modeling/case_study}
In previous work \cite{Stock2023}, we compared the system identification capabilities of \ac{bpinn}, \ac{pinn} and \ac{sindy} under aleatoric uncertainty resulting from noise in the data. This paper focuses on epistemic uncertainty arising from \acp{ibr}.
We first explore the performance using data from the \ac{smib} system. Three different dynamic situations are studied that serve as a baseline for this paper. Second, we implement a 3-bus system with one synchronous generator and two \acp{ibr}. This system aims to support a general understanding of the behavior of all algorithms at different levels of \ac{ibr} penetration. We do not vary the inverter parameters here, thus, the same system with varying shares of synchronous behavior is observed. Third, data from the CIGRE 14-bus system are collected. This system represents a distribution system, with vast penetration of \acp{ibr} and a superordinate transmission system. In this system, we test different parameters for the transmission system and also randomly vary the inverter parameters $J_c$ and $d_c$ by $\pm 20\,\%$ following a uniform distribution. Fourth, we use data from the IEEE 118-bus system. We keep the synchronous generators and inverter parameters fixed and vary the number of inverter-coupled and synchronous generators to achieve different levels of \ac{ibr} penetration.

Note that we estimate one $m$, $d$, $B$ based on a single \cref{eq:regression_problem} for each grid. This process represents the overall system dynamics instead of individual machines. 

\subsection{Network models}
The networks, 3-bus, 14-bus, 118-bus, consist of the corresponding number of nodes $n_n$ nodes and $n_b$ branches.  
The \ac{smib} system is chosen as a baseline as described in the previous section. We use the \ac{smib} in \cref{eq:regression_problem}, so the \ac{smib} regression formulation accurately represents the behavior of the system. We vary the generator parameters as shown in \cref{tab:scenarios}.

\begin{figure}[h!]
    \centering
    \resizebox{0.2\textwidth}{!}{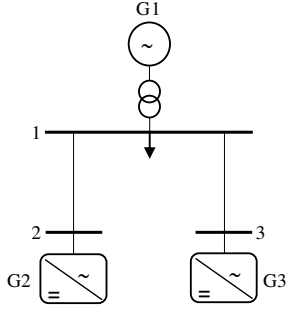}
    \caption{3-bus system with \acp{ibr}}
    \label{fig:3_bus_system}
\end{figure}

The influence of \acp{ibr} is investigated in more detail using a simple 3-bus system. This grid is a consecutive step from the \ac{smib} formulation to a simple system representation that respects the influences of \acp{ibr}. We create a system with one synchronous generator and two \acp{ibr}, as shown in \cref{fig:3_bus_system}. 
Different dynamic situations are simulated by changing the parameters of the synchronous generator as shown in \cref{tab:scenarios}. We still use the same regression formulation \cref{eq:regression_problem} as before, which is no longer accurate due to the \acp{ibr} influence. 

\begin{figure}[h!]
    \centering
    \includegraphics[clip, width=\linewidth]{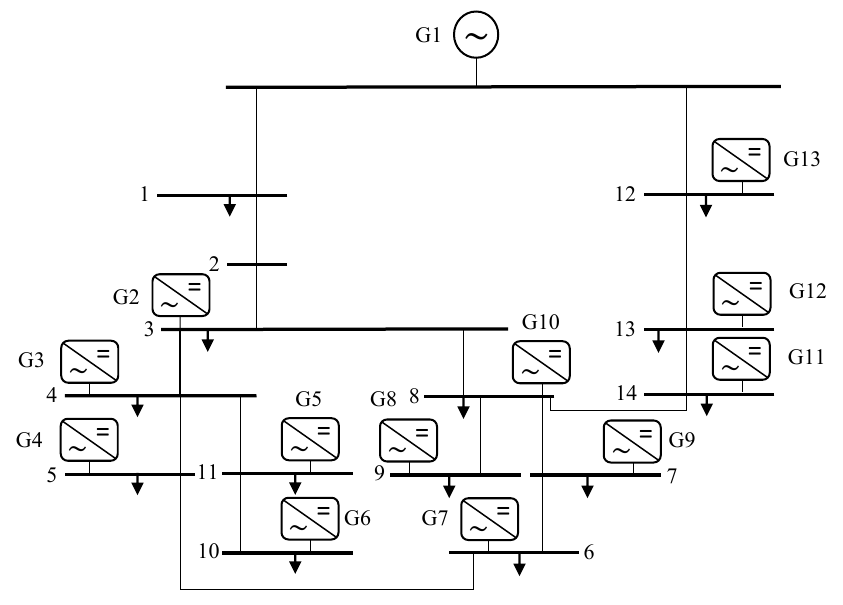}
    \caption{CIGRE 14-bus MV system with \acp{ibr}}
    \label{fig:CIGRE_14_bus_system}
\end{figure}

After exploring the \ac{bpinn} performance on a small-scale system, we perform parameter estimations on the CIGRE 14-bus MV system. This includes 12 \acp{ibr}. The structure of the system is shown in \cref{fig:CIGRE_14_bus_system}.
We vary the dynamics by changing the parameters of the generator, following \cref{tab:scenarios}, and additionally randomly change the $J_c$ and $d_c$ parameters for each of the inverters by $20\,\%$ around the values given in \cref{tbl:Inverter_parameters}.

\begin{figure*}
    \centering
    \includegraphics[clip, width=\linewidth]{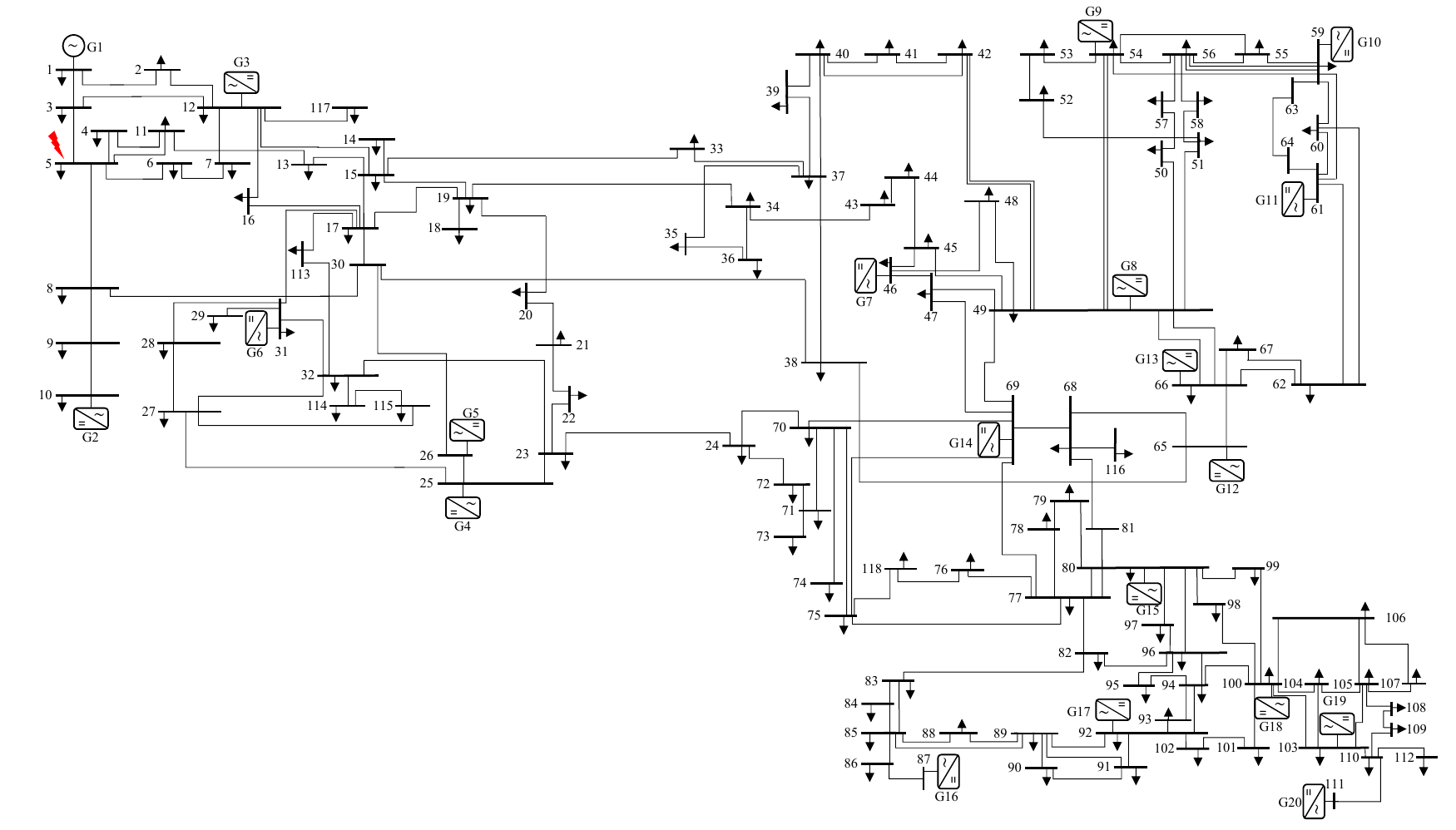}
    \caption{IEEE 118-bus with \acp{ibr}}
    \label{fig:IEEE_118_bus}
\end{figure*}

Finally, we test all algorithms on the IEEE 118-bus system. This system does not have any buses intended for \acp{ibr}, however, we substituted synchronous generators with \acp{ibr}, as illustrated in \cref{fig:IEEE_118_bus}. Different dynamic situations, slow and fast, are created by setting all synchronous generators and \acp{ibr} to fixed parameters and altering the total number of \acp{ibr} and generators. In the fast dynamics scenario, only G1 is a synchronous generator, all other generator buses are equipped with \acp{ibr}. From there on, generator buses G1, G7, G5 are provided with synchronous generators in the medium dynamics scenario. The slow dynamics scenario consists of synchronous generators connected to buses G1, G7, G5, G17, G14, G18, while all other generator buses are equipped with \acp{ibr}.

We start at a stable operating point, so $\dot{\boldsymbol{x}}=0$ at $t=0.0\,\text{s}$ as described in the previous chapter \cref{subsec:sys_ID_eq} and perturb the mechanical power of G1. Measurement data of $\Delta \omega,\delta$ is collected at all synchronous generators.
\begin{table}[!th]
\renewcommand{\arraystretch}{1.2}
\caption{Base evaluation scenarios generator parameters}
\label{tab:scenarios}
\centering
\begin{tabular}{lcc}
\toprule
Scenario & $m_{gen}$ in p.u. & $d_{gen}$ in p.u.\\
\midrule
Fast dynamics & $1.1$ & $0.8$ \\
Medium dynamics & $1.5$ & $1.2$ \\
Slow dynamics & $2.1$ & $1.8$  \\
\bottomrule
\end{tabular}
\end{table}

\subsection{Inverter model}
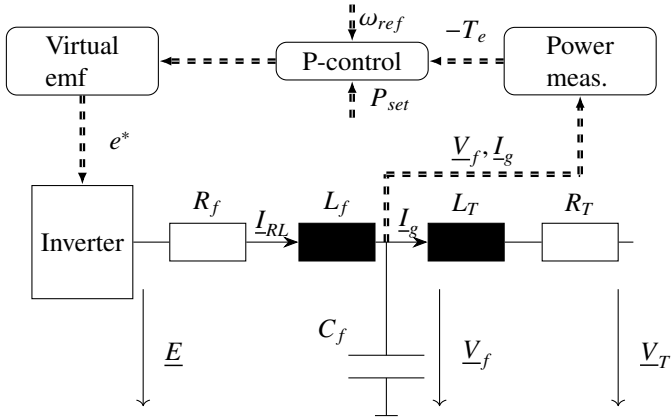
\begin{figure}[h!]
    \centering
    \begin{tikzpicture}[node distance=1.5cm]
        \node (Lf) [text_box,fill=black]{};
        \node (Rf) [text_box, left of=Lf, xshift=-0.2cm]{};
        \node (Lt) [text_box, right of=Lf, fill=black, xshift=0.2cm]{};
        \node (Rt) [text_box, right of=Lt]{};
        \node (meas) [rounded_text_box, above of=Lt,align=left, yshift=0.9cm, xshift=1.5cm]{Power\\meas.};
        \node (Pc) [rounded_text_box, left of=meas, xshift=-1.5cm]{P-control};
        
        \node (inv) [block, left of=Rf, align=left, xshift=-0.15cm]{Inverter};
        \node (emf) [rounded_text_box, above of=inv, align=left, yshift=0.9cm]{Virtual\\emf};

        \coordinate[above of=Pc, yshift=-0.75cm](misc12);
        \coordinate[right of=Lf, xshift=-0.85cm](misc1);
        \coordinate[below of=misc1](misc2);
        \coordinate[left of=misc2, xshift=1cm](misc3);
        \coordinate[right of=misc2, xshift=-1cm](misc4);
        \coordinate[below of=misc2, yshift=1.2cm](misc5);
        \coordinate[left of=misc5, xshift=1cm](misc6);
        \coordinate[right of=misc5, xshift=-1cm](misc7);
        \coordinate[below of=misc5, yshift=1cm](misc8);
        \coordinate[left of=misc8, xshift=1.35cm](misc9);
        \coordinate[right of=misc8, xshift=-1.35cm](misc10);
        \coordinate[right of=Rt, xshift=-0.8cm](misc11);

        \coordinate[above of=misc1, yshift=-0.6cm](misc13);
        \coordinate[below of=meas](misc14);

        \node[above of=Rt, align=left, yshift=-1cm]{$R_T$};
        \node[above of=Lt, align=left, yshift=-1cm]{$L_T$};
        \node[above of=Rf, align=left, yshift=-1cm]{$R_f$};
        \node[above of=Lf, align=left, yshift=-1cm]{$L_f$};
        \node[above of=misc2, align=left, yshift=-1.2cm, xshift=-0.7cm]{$C_f$};

        \node(upe)[right of=inv, align=left, xshift=-0.7cm, yshift=-0.5cm]{};
        \node(loe)[right of=inv, align=left, xshift=-0.7cm, yshift=-2.3cm]{};
        \draw[->](upe)--(loe){};
        \node[right of=inv, align=left, xshift=-0.3cm, yshift=-1.5cm]{$\underline{E}$};

        \node(uVf)[right of=misc1, align=left, xshift=-0.8cm, yshift=-0.5cm]{};
        \node(lVf)[right of=misc1, align=left, xshift=-0.8cm, yshift=-2.3cm]{};
        \draw[->](uVf)--(lVf){};
        \node[right of=misc1, align=left, xshift=-0.3cm, yshift=-1.5cm]{$\underline{V}_f$};
        \node(uVt)[right of=Rt, align=left, xshift=-1cm, yshift=-0.5cm]{};
        \node(lVt)[right of=Rt, align=left, xshift=-1cm, yshift=-2.3cm]{};
        \draw[->](uVt)--(lVt){};
        \node[right of=Rt, align=left, xshift=-0.5cm, yshift=-1.5cm]{$\underline{V}_T$};

        \draw[dashed, line width=1pt, double distance=0.1pt](misc1)--(misc13);
        \draw[dashed, line width=1pt, double distance=0.1pt](misc13)--(misc14)node[midway, align=center, yshift=0.3cm]{$\underline{V}_f, \underline{I}_g$};
        \draw[dashed, line width=1pt, double distance=0.1pt,
             arrows = {-Latex[length=0pt 3 .5]}](misc14)--(meas);
        \draw[dashed, line width=1pt, double distance=0.1pt,
             arrows = {-Latex[length=0pt 3 .5]}](misc12)--(Pc)node[midway, align=center, xshift=0.4cm]{$\omega_{ref}$};
        \coordinate[below of=Pc, yshift=0.75cm](misc22);
        \draw[dashed, line width=1pt, double distance=0.1pt,
             arrows = {-Latex[length=0pt 3 .5]}](misc22)--(Pc)node[midway, align=center, xshift=0.5cm]{$P_{set}$};
        \draw[](misc3)--(misc4);
        \draw[](misc1)--(misc2);
        \draw[](misc6)--(misc7);
        \draw[](misc5)--(misc8);
        \draw[](misc9)--(misc10);
        \draw[](inv)--(Rf);
        \draw[arrow](Rf)--(Lf)node[midway, align=center, yshift=0.25cm]{$\underline{I}_{RL}$};
        \draw[arrow](Lf)--(Lt)node[midway, align=center, xshift=0.1cm,yshift=0.25cm]{$\underline{I}_g$};
        \draw[](Lt)--(Rt);
        \draw[](Rt)--(misc11);
        \draw[ dashed, line width=1pt, double distance=0.1pt,
             arrows = {-Latex[length=0pt 3 .5]}](emf)--(inv) node[midway, align=center, xshift=0.5cm]{$e^*$};
        \draw[ dashed, line width=1pt, double distance=0.1pt,
             arrows = {-Latex[length=0pt 3 .5]}](Pc)--(emf);
        \draw[ dashed, line width=1pt, double distance=0.1pt,
             arrows = {-Latex[length=0pt 3 .5]}](meas)--(Pc)node[midway, align=center,yshift=0.4cm]{$-T_e$};


    \end{tikzpicture}
    \caption{Simplified inverter diagram}
    \label{fig:Inverter_1_line_diagram}
\end{figure}

We model \acp{ibr} as a battery connected via a synchronverter \cite{Zhong2011, Rosso2017}. A simplified diagram of the electrical and control parts is shown in \cref{fig:Inverter_1_line_diagram}. 

For simplification, we neglect the DC side and switching, and focus solely on calculation of the synchronverter control, coupled with an RLC filter, $R_f, L_f, C_f$, and transformer, $R_T, L_T$. The power side of the inverter is modelled as follows
\begin{align}
    \dot{\underline{I}}_{RL}&=\frac{1}{L_f}(\underline{E}-\underline{V}_f-R_f \underline{I}_{RL}) \\
    \dot{\underline{V}}_f&=\frac{1}{C_f}(\underline{I}_{RL}-\underline{I}_g)\\
    \dot{\underline{I}}_g&=\frac{1}{L_t}(\underline{V}_f-\underline{V}_T-R_T \underline{I}_g)
\end{align}
The active power control side is determined in the synchronverter topology, which imitates the swing equation by calculating a virtual angular frequency $\omega_c$ and a virtual angle $\delta_c$: 
\begin{align}
    \dot{\omega}_c&= \frac{1}{J_c}(-d_c (\omega_{ref}-\omega_c)-T_e+T_m)\\
    \dot{\delta}_c&= \omega_c\\
    \dot{e}^*&=\delta_c M_f i_f \text{sin}(\delta_c).
\end{align}
The mechanical torque $T_m$ is based on a power setpoint $P_{set}$, which is provided during normal operation. We only utilize the inertial response of the batteries, hence, $P_{set}=0$.

The inverter parameters are presented in \cref{tbl:Inverter_parameters}.
\begin{table}[!th]
\renewcommand{\arraystretch}{1.0}
\caption{Synchronverter parameters}
\label{tbl:Inverter_parameters}
\centering
\resizebox{0.8\linewidth}{!}{%
\begin{tabular}{llr}
\toprule
Description & Symbol & Value\\
\midrule
Filter resistance& $R_f$ & $0.375\,\text{m}\Omega$\\
Filter inductance & $L_f$ & $0.3\,$mH\\
Filter capacitance & $C_f$ & $0.25\,\text{mF}$\\
Transformer resistance & $R_T$ & $0.22\,\text{m}\Omega$\\
Transformer inductance & $L_T $& $0.3\,\text{mH}$\\
 Virtual Inertia & $J_c$ & $4.052\cdot 10^{-4}$\\
 Virtual Damping & $d_c$ & $0.679$\\
\bottomrule
\end{tabular}}
\end{table}

We also added a frequency deadband to the inverter controller. It avoids taking actions when the deviation is too small.  

\subsection{Synchronous generator model}
The synchronous generator is represented by a third order system, that models the dynamic behavior of frequency deviation $\Delta \omega_{gen}$ and angle $\delta_{gen}$ based on the swing equation and the dynamics of governor control \cite{Bergen1981} 
\begin{align}
    \dot{\delta}_{gen}&=\Delta\omega_{gen}\\
    \Delta\dot{\omega}_{gen}&=\frac{1}{m_{gen,k}}(-d_{gen,k}\Delta\omega_{gen}-\sum_j B_{kj}\text{sin}(\delta_j-\delta_k)\\
    \nonumber&+P_{m,k}+P_{gov,k})\\
    \dot{P}_{gov,k}&=-\frac{1}{T_s}(\Delta\omega_{gen}+P_{gov,k}).
\end{align}
$m_{gen,k}$ and $d_{gen,k}$ are the inertia and damping of the generator and $T_s$ the governor time constant.

\section{Results}
\label{sec:results}
We compare performance by assessing the \ac{mape} of the measurement trajectories $\boldsymbol{x}$ and the reconstruction $\hat{\boldsymbol{x}}(\hat{\boldsymbol{\lambda}})$ based on the estimated system parameters $\hat{\boldsymbol{\lambda}}$. The \ac{mape} is defined as follows: 
\begin{align}
    \text{MAPE}=100\,\%\cdot \frac{1}{n_z}\sum_i^{n_z}\left|\frac{x_i-\hat{x}_{i}(\hat{\lambda})}{x_i}\right|.
    \label{eq:MAPE}
\end{align}

\begin{figure}
    \centering
    \begin{subfigure}[T]{0.48\linewidth}
        \centering
        \hspace{0.8cm}$\text{MAPE}_{\Delta\omega}$ in $\%$
\begin{tikzpicture}

\definecolor{darkgray176}{RGB}{176,176,176}
\definecolor{lightgray204}{RGB}{204,204,204}
\definecolor{deepblue}{RGB}{21,76,121}

\begin{axis}[width=1.7in,
                view={0}{90},
                unbounded coords=jump,
                colormap/cool,
                point meta min=0,
                point meta max=50,
                tick align=outside,
                tick pos=left,
                x grid style={darkgray176},
                xlabel={$\kappa_{prior}$},
                xtick style={color=black},
                y grid style={darkgray176},
                ylabel={$\mu_{ prior}$},
                ytick style={color=black},
                anchor=north,
                yshift=-2cm,
                axis lines = left,
                ]
\addplot3 [surf, mesh/ordering=y varies,mesh/rows=10,mesh/cols=11]
    table{
    x  y color
	5.0	0.0	2.349270453190822
	10.0	0.0	2.3469089376073367
	15.0	0.0	2.22993522108692
	20.0	0.0	2.172886456083879
	25.0	0.0	7.523693846098088
	30.0	0.0	29.663640002775203
	35.0	0.0	52.90904272730527
	40.0	0.0	45.95191903460103
	45.0	0.0	41.981512757282744
	50.0	0.0	40.70257033854135
	5.0	1.0	2.3523715691222096
	10.0	1.0	2.362330910589392
	15.0	1.0	2.354277163480838
	20.0	1.0	2.2630540194778277
	25.0	1.0	2.126928955014823
	30.0	1.0	5.454863421985789
	35.0	1.0	24.350115677655808
	40.0	1.0	46.950647367072314
	45.0	1.0	43.66365632579141
50.0	1.0	39.16708252516128
	5.0	2.0	2.3520462381187173
	10.0	2.0	2.3611743009273307
	15.0	2.0	2.3651078099550404
	20.0	2.0	2.358527339251424
	25.0	2.0	2.279464965023308
	30.0	2.0	2.1125958195610766
	35.0	2.0	4.2102477638119264
	40.0	2.0	19.25536417942799
	45.0	2.0	40.68077136610819
	50.0	2.0	42.832943757658015
	5.0	3.0	2.3505292877994983
	10.0	3.0	2.357842237235055
	15.0	3.0	2.363163172723321
	20.0	3.0	2.3640531301138945
	25.0	3.0	2.350274416807879
	30.0	3.0	2.280124512456186
	35.0	3.0	2.084344768591349
	40.0	3.0	2.9754915983794294
	45.0	3.0	15.461011285566025
	50.0	3.0	34.975025829062695
	5.0	4.0	2.346039329964892
	10.0	4.0	2.359497462351672
	15.0	4.0	2.3597525361314005
	20.0	4.0	2.3575559439782805
	25.0	4.0	2.347988363962769
	30.0	4.0	2.31746897614361
	35.0	4.0	2.2361131539710155
	40.0	4.0	2.012869876415734
	45.0	4.0	2.4349921395915137
	50.0	4.0	13.222540685169331
	5.0	5.0	2.3432378186776517
	10.0	5.0	2.3563796189816815
	15.0	5.0	2.3593840795268806
	20.0	5.0	2.35912339661699
	25.0	5.0	2.3342641535935216
	30.0	5.0	2.2888615270487045
	35.0	5.0	2.2017371390208793
	40.0	5.0	2.073361245707074
	45.0	5.0	1.9542250357241404
	50.0	5.0	3.715419567339747
	5.0	6.0	2.3268753002561846
	10.0	6.0	2.3340434171709874
	15.0	6.0	2.3520583536621125
	20.0	6.0	2.360284684695602
	25.0	6.0	2.352530747367656
	30.0	6.0	2.260199650320427
	35.0	6.0	2.0815648761341636
	40.0	6.0	2.1360425957209044
	45.0	6.0	3.020474901397026
	50.0	6.0	4.38979070169337
	5.0	7.0	2.268010813304662
	10.0	7.0	2.259142375291048
	15.0	7.0	2.277862170290169
	20.0	7.0	2.3176619425369105
	25.0	7.0	2.332167662653225
	30.0	7.0	2.308809960805324
	35.0	7.0	2.0787598300214283
	40.0	7.0	3.62454591030858
	45.0	7.0	5.907132552318588
	50.0	7.0	7.933929086533066
	5.0	8.0	2.550219159054955
	10.0	8.0	2.064742492164786
	15.0	8.0	2.0345657728470052
	20.0	8.0	2.063238424816818
	25.0	8.0	2.1129147823173233
	30.0	8.0	2.1706462453917106
	35.0	8.0	2.134334607714734
	40.0	8.0	2.455373575212697
	45.0	8.0	8.197363831129598
	50.0	8.0	11.676260711998163
	5.0	9.0	8.916724262367346
	10.0	9.0	5.350846659073903
	15.0	9.0	5.310192429887392
	20.0	9.0	4.932364073936733
	25.0	9.0	4.700750794329494
	30.0	9.0	3.907546983158832
	35.0	9.0	2.9286656539557843
	40.0	9.0	3.239575403027714
	45.0	9.0	5.519957196153998
	50.0	9.0	13.086734814912102
5.0	10.0	25.407354407287553
	10.0	10.0	13.835847239296434
	15.0	10.0	11.599429084442967
	20.0	10.0	12.244948812481876
	25.0	10.0	12.139832041612522
	30.0	10.0	11.199637903094379
	35.0	10.0	10.265089224331552
	40.0	10.0	8.942660508469132
	45.0	10.0	8.892614364882505
	50.0	10.0	10.85023313471418

};

\end{axis}

\end{tikzpicture}
        \label{fig:prior_sensitivity_omega}
    \end{subfigure}
   \hfill
    \begin{subfigure}[T]{0.48\linewidth}
        \centering
        $\text{MAPE}_{\delta}$ in $\%$
    
\begin{tikzpicture}

\definecolor{darkgray176}{RGB}{176,176,176}
\definecolor{lightgray204}{RGB}{204,204,204}
\definecolor{deepblue}{RGB}{21,76,121}

\begin{axis}[width=1.7in,
                view={0}{90},
                unbounded coords=jump,
                colorbar,
                colormap/cool,
                point meta min=0,
                point meta max=55,
                colorbar style={width=0.2cm,
                height=1.8cm},
                tick align=outside,
                tick pos=left,
                x grid style={darkgray176},
                xlabel={$\kappa_{prior}$},
                xtick style={color=black},
                y grid style={darkgray176},
                ytick style={color=black},
                anchor=north,
                yshift=-2cm,
                axis lines = left,
                ]
\addplot3 [surf, mesh/ordering=y varies,mesh/rows=10,mesh/cols=11]
    table{
	x	y	omega
	5.0	0.0	1.8229668580408198
	10.0	0.0	1.8216071459816037
	15.0	0.0	1.6269309412328312
	20.0	0.0	1.5245033602716989
	25.0	0.0	3.2790074458133063
	30.0	0.0	13.531400331692279
	35.0	0.0	22.772009944461004
	40.0	0.0	25.65094206839329
	45.0	0.0	29.081612681941593
	50.0	0.0	31.098004184402033
	5.0	1.0	1.8284944544827542
	10.0	1.0	1.8482571705723143
	15.0	1.0	1.8345933451781204
	20.0	1.0	1.686289834124643
	25.0	1.0	1.471228642301665
	30.0	1.0	2.2958076091412103
	35.0	1.0	11.24777086671626
	40.0	1.0	21.189121690635606
	45.0	1.0	27.251942748824725
	50.0	1.0	34.996617579106974
	5.0	2.0	1.8278342235591856
	10.0	2.0	1.8456542433757959
	15.0	2.0	1.853592945680312
	20.0	2.0	1.8421804200918042
	25.0	2.0	1.7089311364090962
	30.0	2.0	1.4083826002454427
	35.0	2.0	1.7876763981585189
	40.0	2.0	8.976823928262561
	45.0	2.0	19.390666707379197
	50.0	2.0	28.06310187611036
	5.0	3.0	1.8247232618691305
	10.0	3.0	1.83901505552758
	15.0	3.0	1.8494655569725078
	20.0	3.0	1.8513302518697896
	25.0	3.0	1.8257077396653472
	30.0	3.0	1.701735083000833
	35.0	3.0	1.306635448324523
	40.0	3.0	1.1605353148724038
	45.0	3.0	7.362815106359665
	50.0	3.0	18.14968770864892
	5.0	4.0	1.8158676772084972
	10.0	4.0	1.8424697144173177
	15.0	4.0	1.8425205007122452
	20.0	4.0	1.8383423162851378
	25.0	4.0	1.8196887824817813
	30.0	4.0	1.7617012455295276
	35.0	4.0	1.6075799756907774
	40.0	4.0	1.1121667375647974
	45.0	4.0	0.5493462773339708
	50.0	4.0	7.014003267759186
	5.0	5.0	1.8107574736991126
	10.0	5.0	1.8363686770572323
	15.0	5.0	1.8420505891853487
	20.0	5.0	1.8418750170665281
	25.0	5.0	1.7925929339281566
	30.0	5.0	1.7042941181923856
	35.0	5.0	1.5353538846015036
	40.0	5.0	1.2795006129780941
	45.0	5.0	0.908897007386619
	50.0	5.0	2.3582353479896248
	5.0	6.0	1.7792973015092621
	10.0	6.0	1.7933366270857687
	15.0	6.0	1.8280797660704715
	20.0	6.0	1.84391475033063
	25.0	6.0	1.8282748657143226
	30.0	6.0	1.6494864728742418
	35.0	6.0	1.29939536621952
	40.0	6.0	1.338811272047137
	45.0	6.0	2.6678849196300702
	50.0	6.0	4.552685612673234
	5.0	7.0	1.6578153048136786
	10.0	7.0	1.6487936923686757
	15.0	7.0	1.6843568087339944
	20.0	7.0	1.761319120031969
	25.0	7.0	1.788674949602432
	30.0	7.0	1.743196202133367
	35.0	7.0	1.2939653761617915
	40.0	7.0	3.626251698838976
	45.0	7.0	7.04089765970867
	50.0	7.0	10.005689170943082
	5.0	8.0	1.9990431020610415
	10.0	8.0	1.2630497953882636
	15.0	8.0	1.1996462085430752
	20.0	8.0	1.2622508165389543
	25.0	8.0	1.3626781292073695
	30.0	8.0	1.4746329644562657
	35.0	8.0	1.4040873584928182
	40.0	8.0	1.83738747382622
	45.0	8.0	10.463721698460452
	50.0	8.0	15.637575976975961
	5.0	9.0	9.66432925219042
	10.0	9.0	6.62225557893391
	15.0	9.0	6.167588352149796
	20.0	9.0	5.540901343204256
	25.0	9.0	5.220031727135085
	30.0	9.0	4.048558579319228
	35.0	9.0	2.564939106317625
	40.0	9.0	3.042128454222892
	45.0	9.0	6.479404602671453
	50.0	9.0	17.77151888339878
	5.0	10.0	26.829546293792454
    10.0	10.0	18.774425704680382
	15.0	10.0	16.19075984298627
	20.0	10.0	16.479770091454814
	25.0	10.0	16.226745488789632
	30.0	10.0	14.873842705658666
	35.0	10.0	13.519382474561978
	40.0	10.0	11.567268816711387
	45.0	10.0	11.498774473689338
	50.0	10.0	14.419187122832971

};

\end{axis}

\end{tikzpicture}
    \label{fig:prior_sensitivity_delta}
    \end{subfigure}
    \caption{MAPE sensitivity to $\boldsymbol{\lambda}$ prior}
    \label{fig:prior_influence}
\end{figure}
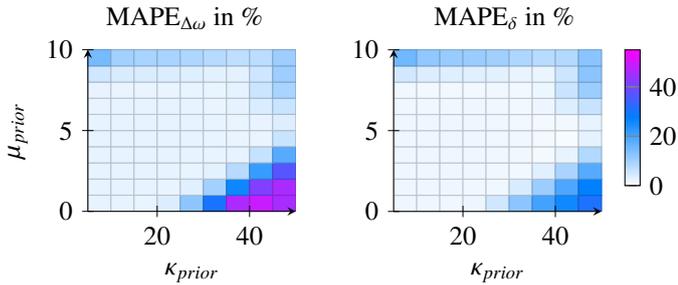

We evaluate the sensitivity of the parameter estimation error to weakly-informative prior design in the following based on \cref{eq:priors_for_lambda}. The aim is to balance the amount of information in the prior and avoid non-informative priors. A totally non-informative prior would bypass the Bayesian approach, while a prior containing too much information would cause a biased estimate. 
In general, a similar sensitivity of the \acp{mape} can be found with respect to $\kappa_{prior}$ and $\mu_{prior}$ for $\Delta \omega$ and $\delta$ in \cref{fig:prior_influence}. When $\mu_{prior}$ is small, i.e. range of approximately $[0,5]$, small estimation errors are achieved. A similar behavior is found for $\kappa_{prior}$, which constantly achieves low errors in the range of up to 25 for both quantities. Both behaviors are more pronounced for $\text{MAPE}_{\Delta\omega}$ than for $\text{MAPE}_{\delta}$. High $\mu_{prior}$ and $\kappa_{prior}$ lead to less informative priors. This is indicated by the decreased density in \cref{fig:example_normal_gamma_distribution}. \cref{fig:prior_influence} reveals that too little information in the prior result in a larger error. 
Based on these results, both parameters seem to be best set in the medium region around $\mu_{prior}=[1,4]$ and $\kappa_{prior}=[10,20]$.
For this paper, we use $\mu_{{prior}}=1.0$ and $\kappa_{{prior}}=25.0$ in all test scenarios.

\subsection{Influence of epistemic uncertainty}
\begin{table*}
\renewcommand{\arraystretch}{1.0}
\caption{Influence of epistemic uncertainty in 3-Bus, 14-bus and 118-bus system and different dynamic scenarios}
\label{tbl:Noise_results}
\centering
\resizebox{\textwidth}{!}{
\begin{tabular}{ll|rrrrrrrrrrrr}
\toprule
& &  \multicolumn{4}{c}{Fast dynamics} & \multicolumn{4}{c}{Medium dynamics} & \multicolumn{4}{c}{Slow dynamics} \\
System & Algorithm & $\text{MAPE}_{\Delta \omega}\,[\%]$ & $2\sigma\,[\%]$& $\text{MAPE}_{\delta} \,[\%]$ &$2\sigma\,[\%]$& $\text{MAPE}_{\Delta \omega}\,[\%]$ & $2\sigma\,[\%]$& $\text{MAPE}_{\delta} \,[\%]$ &$2\sigma\,[\%]$ & $\text{MAPE}_{\Delta \omega}\,[\%]$ & $2\sigma\,[\%]$& $\text{MAPE}_{\delta} \,[\%]$ &$2\sigma\,[\%]$\\
\midrule

\multirow{3}{*}{SMIB}
 & BPINN &  1.320 &  7.808 &  0.716 & 11.221 &  1.794 & 12.158 &  1.055 & 16.744 &  2.118 & 15.290 &  1.334 & 20.283\\
&PINN&   0.273 &  &   0.284 &  &   0.247 &  &   0.288 &  &   2.821 &  &   3.940 &  \\
& SINDy &  0.013 &   &  0.015 &   &  0.023 &   &  0.025 &   &  0.038 &   &  0.040 &     \\
\hline
\multirow{3}{*}{3-bus} & BPINN &   9.828 &  12.007 &   5.576 &   9.365 &   5.672&   5.715 &   3.888 &   4.009 &   2.127 &  10.617 &   1.471 &   9.907\\
&PINN&   18.295 &  &  12.475 &  &   9.111 &  &   6.196 &  &   5.009 &  &   4.245 &  \\
& SINDy &  56.972 &    &  55.949 &    &  60.107 &    &  59.821 &    &  43.957 &    &  43.932 &   \\
\hline
\multirow{3}{*}{14-bus} &BPINN &   7.785 &   7.651 &   4.682 &   4.312 &   4.077 &   6.465 &   2.933 &   5.314 &   1.457 &  12.816 &   1.033 &  12.063  \\
&PINN&   14.231 &  &   9.741 &  &   6.824 &  &   5.016 &  &   3.498 &  &   3.199 &   \\
& SINDy &  50.509 &    &  49.955 &    &  47.961 &    &  47.853 &    &  30.723 &    &  30.726 &     \\
\hline
\multirow{3}{*}{118-bus} &BPINN &   1.079 &  13.543 &   0.832 &  12.308 &   0.496 &  14.510 &   0.379 &  13.310 &   0.506 &  14.643 &   0.374 &  13.418\\
&PINN& 3.699 &  &   2.545 &  &   3.972 &  &   2.692 &  &   4.487 &  &   3.026&  \\
& SINDy &  23.174 &    &  24.624 &    &  12.891 &    &  13.941 &    &  10.862 &    &  11.933 &    \\

\bottomrule
\end{tabular}
}
\end{table*}

In this section, we compare the \ac{bpinn} for system identification under model uncertainty on four different systems for three dynamic settings each and contrast the performance with \ac{sindy} and \ac{pinn}. The results are presented in \cref{tbl:Noise_results}. 
These three approaches, \ac{bpinn}, \ac{pinn} and \ac{sindy} are substantially different. \ac{sindy} uses a point-wise method to obtain the derivatives of the states $\boldsymbol{x}$ while \ac{pinn} and \ac{bpinn} apply automatic differentiation. The latter technique calculates the derivatives with respect to inputs $u$ by dismantling the surrogate model, i.e. the \ac{bnn}, into primitives with known derivatives. These are combined using the chain rule based on a computational graph. When the surrogate model is accurate, the derivative will also be accurate, ultimately leading to precise system parameter estimates.
In addition, the \ac{bpinn} provides richer information through the posterior standard deviation, which is not given by either \ac{sindy} or \ac{pinn}. In this paper, we provide the reconstruction error \ac{mape} for each state $\boldsymbol{x}$ and also give the corresponding posterior standard deviation $2\sigma$ from the \ac{bpinn}. 

The results demonstrate that \ac{sindy} achieves lower errors than \ac{bpinn} and also \ac{pinn} for the \ac{smib} system estimation. The data utilized accurately represent the formulation of the regression problem, that is, \cref{eq:regression_problem}, thus, \ac{sindy}s point-wise approach is advantageous here. On the contrary, the \ac{bpinn} is unable to achieve lower errors, since it fits a family of surrogate models. The determined distribution can only be narrowed down to a certain point with confidence. 
It also becomes apparent that the posterior standard deviation becomes wider with decreasing dynamics, i.e. from fast to slow. This stems from the sensitivity of the measurement trajectories $\mathcal{D}$ to changes in the system parameters $\boldsymbol{\lambda}$. Slower dynamics tend to lead to less sensitive state trajectories $\boldsymbol{x}$ considering a change in parameters $\boldsymbol{\lambda}$. This decreases the \ac{bpinn}s confidence, since there are fewer contradictory values in the initial distribution. A similar behavior can be found for all test grids: the fastest dynamics always result in the most narrow distribution compared to slower dynamics. For the same reason, we find larger errors in slower dynamics for the \ac{pinn}. 
The errors $\text{MAPE}_{\Delta \omega}$ are comparable to the $\text{MAPE}_{\delta}$ errors for \ac{sindy} in \ac{smib}. The \ac{bpinn} produces slightly larger errors for $\text{MAPE}_{\Delta \omega}$. A similar behavior is found for the \ac{pinn}. 

The previous paragraph evaluates the \ac{smib} benchmark results. In the following, we introduce epistemic uncertainty in the three other grids that incorporate \acp{ibr}. The results show that the \ac{bpinn} is able to achieve significantly lower errors than \ac{sindy} and also the \ac{pinn} in all cases. Most times, the \ac{bpinn} error is smaller by factor ten compared to \ac{sindy} error, in some cases it is even close to factor 90. 
This stems from the fact that \ac{sindy}s estimation approach is not beneficial anymore, since the regression problem, \cref{eq:regression_problem}, does not describe the power system behavior to the full extent. \ac{sindy} often determines a parameter set that only partially represents the system behavior. The \ac{pinn} and \ac{bpinn} show better performance. 
Both seek to obtain a surrogate model which enables state prediction for any point in time. The system parameters $\hat{\boldsymbol{\lambda}}$ are calculated based on these state estimates and the corresponding derivatives. This averages out underlying effects, since \ac{bpinn} and \ac{pinn} use automatic differentiation compared to point-wise differentiation. 
The \ac{bpinn} achieves better performance compared to the \ac{pinn} due to its Bayesian nature, which aims to exclude contradictory parameters from the posterior distribution. Higher uncertainty causes a distribution that consists of widespread parameters, thus, it most likely covers the correct solution. The \ac{bpinn} and \ac{pinn} are different by approximately factor two to three in most cases. 

For all algorithms, the error decreases with slower dynamics for the majority of dynamic settings. This stems from the fact that the fast \ac{ibr} dynamics vanish in the dynamics of the overall grid. In faster dynamic scenarios, the \ac{ibr} dynamics are more pronounced and dominate the system behavior. This effect can be distinctively observed in the 14-bus system, where \ac{sindy} achieves high errors in fast dynamics and significantly lower errors in slow dynamics. The 14-bus system includes a large number of \acp{ibr} compared to its size. Similar behaviors can be found for the \ac{bpinn} and \ac{pinn}.  

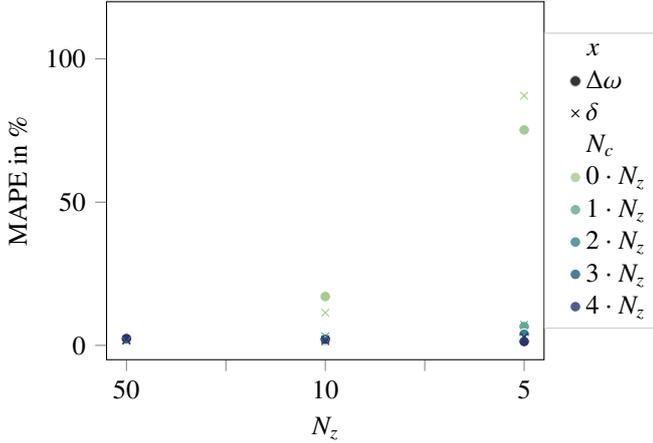
\begin{figure}
    \centering
\begin{tikzpicture}

\definecolor{cadetblue96170144}{RGB}{96,170,144}
\definecolor{darkgray176}{RGB}{176,176,176}
\definecolor{darkseagreen164204144}{RGB}{164,204,144}
\definecolor{lightgray204}{RGB}{204,204,204}
\definecolor{midnightblue4448113}{RGB}{44,48,113}
\definecolor{steelblue51132141}{RGB}{51,132,141}
\definecolor{teal2993134}{RGB}{29,93,134}

\begin{axis}[
legend cell align={left},
legend style={
  fill opacity=0.8,
  draw opacity=1,
  text opacity=1,
  at={(1,0.5)},
  anchor=west,
  draw=lightgray204
},
width=.4\textwidth,
tick align=outside,
tick pos=left,
x grid style={darkgray176},
xlabel={\(\displaystyle N_z\)},
xticklabels={, 50,, 10, ,5},
xmin=-0.1, xmax=2.1,
y grid style={darkgray176},
ylabel={MAPE in \(\displaystyle \%\)},
ymin=-5, ymax=120,
name=first axis,
anchor=south,
]
\addlegendimage{empty legend}\addlegendentry{$x$}
\addlegendimage{mark=*, only marks}\addlegendentry{$\Delta \omega$}
\addlegendimage{draw=black, mark=x, only marks}\addlegendentry{$\delta$}
\addlegendimage{empty legend}\addlegendentry{$N_c$}
\addplot [ draw=white, fill=darkseagreen164204144, mark=*, only marks]
coordinates{(0.0, 2.539553413897916) (1.0,17.062062782935367)(2.0, 75.20860258974409)
};

\addlegendentry{$0 \cdot N_z$}
\addplot [draw=white, fill=cadetblue96170144, mark=*, only marks]
coordinates{(0.0, 2.2172191643996335) (1.0, 2.5003246876205294)(2.0, 6.568881472690627)};

\addlegendentry{$1\cdot N_z$}
\addplot [draw=white, fill=steelblue51132141, mark=*, only marks]
coordinates{(0.0, 2.080429354133752) (1.0, 1.6555570800314745)(2.0, 3.916760314965428)};

\addlegendentry{$2\cdot N_z$}
\addplot [draw=white, fill=teal2993134, mark=*, only marks]
coordinates{(0.0, 2.4345894605601957) (1.0, 2.1114540726876423) (2.0,1.7176416330796251)
};

\addlegendentry{$3\cdot N_z$}
\addplot [draw=white, fill=midnightblue4448113, mark=*, only marks]
coordinates{
(0.0, 2.3837283786839643) (1.0, 2.13971764880614)  (2.0,1.3420808235236608 )};

\addlegendentry{$4\cdot N_z$}

\addplot [ draw=darkseagreen164204144, fill=darkseagreen164204144, mark=x, only marks]
coordinates{(0.0, 1.6838688664087722) (1.0, 11.429923169762578)(2.0, 87.16373566429816)
};

\addplot [draw=cadetblue96170144, fill=cadetblue96170144, mark=x, only marks]
coordinates{(0.0, 1.5770558806066963) (1.0, 3.163311305696538)(2.0, 7.234229282580676	)};

\addplot [draw=steelblue51132141, fill=steelblue51132141, mark=x, only marks]
coordinates{(0.0, 1.8904646247230337) (1.0, 2.0400352801789365)(2.0, 3.0910304289424437)};

\addplot [draw=teal2993134, fill=teal2993134, mark=x, only marks]
coordinates{(0.0, 1.6896371128145018) (1.0, 2.0266143616942176) (2.0,3.258325207536422)};

\addplot [draw=midnightblue4448113, fill=midnightblue4448113, mark=x, only marks]
coordinates{(0.0, 2.024932669432593	) (1.0, 1.4578125933122064)  (2.0,2.8768070427943524 )};

\end{axis}

\end{tikzpicture}
    \caption{Influence of collocations points ($N_c$) and number of samples ($N_z$) on the estimation accuracy (IEEE 118-bus system "fast dynamics")}
    \label{fig:mape_step_size_coll}
\end{figure}

\paragraph{Influence of sampling frequency and collocation points}
\cref{sec:methodology} describes the \ac{bpinn} and \ac{pinn} ability to generate additional data points, collocation points $N_c$, which can potentially improve the training performance. These collocation points augment the dataset and can significantly reduce the estimation error in case of sparse data. In this paragraph, we explore the influence of collocation points $N_c$ and the sampling frequency on the estimation accuracy. The fast dynamics scenario and the 118-bus system serve as a guiding example in \cref{fig:mape_step_size_coll}. 
The results show that a small number of measurement samples $N_z$ substantially increases the estimation error up to $88\,\%$. In that case, the amount of samples does not allow to create a solid trajectory and system parameter estimate. The estimation of system parameters $\hat{\boldsymbol{\lambda}}$ is based on uncertain estimates $\hat{\boldsymbol{x}}$, which leads to inaccurate overall results. To address the lack of data, supplementary data points can be generated to assess the physics loss, i.e. the collocation points $N_c$. \cref{fig:mape_step_size_coll} reveals, that a small number of collocation points, $N_c=1\cdot N_z$, already reduces the \ac{mape} for $N_z=10$ close to a level comparable to $N_z=50$. However, $N_c=2\cdot N_z$ is required to reach the target error. 
We find a similar behavior for $\text{MAPE}_{\Delta \omega}$ and $\text{MAPE}_{\delta}$. Adding more collocation points to the training data does not significantly reduce the error when $N_z=10$. However, a dataset of $N_z=5$ requires $N_c=4\cdot N_z$ to achieve an error comparable to $N_z=50$.

\subsection{Transfer learning}
The general idea behind transfer learning arises from human learning, which often uses previously learned knowledge to solve new or similar tasks in another domain. Most \ac{ml} methods however assume that the training and later on operational domain are the same, which in reality is not true for most cases. This often requires a comprehensive retraining or even rebuilding of the model when the feature or domain space changes \cite{PAN2010}, resulting in manifold problems. First, it can be computationally expensive, second, the data gathering or labeling can take a long time, or even be impossible for the new domain or feature space. Therefore, it would be beneficial for the \ac{ml} algorithm to reduce the need for training steps or training data from the new domain or feature space. This also holds for the \ac{bpinn}

In this section, we seek to reduce the required amount of training iterations and data through transfer learning. To achieve this, we pretrain on a \ac{smib} with 1000 iterations and transfer the learned behavior to a larger system, the 118-bus system. 
According to the previous introduction, our learning task will remain the same, but the learning domain changes. 
We strive to explore two different questions in this section: First, how many training iterations are required to achieve a comparable estimation result with pretraining compared to exclusive training in target domain? Second, can we reduce the amount of data required by pretraining on the \ac{smib}?

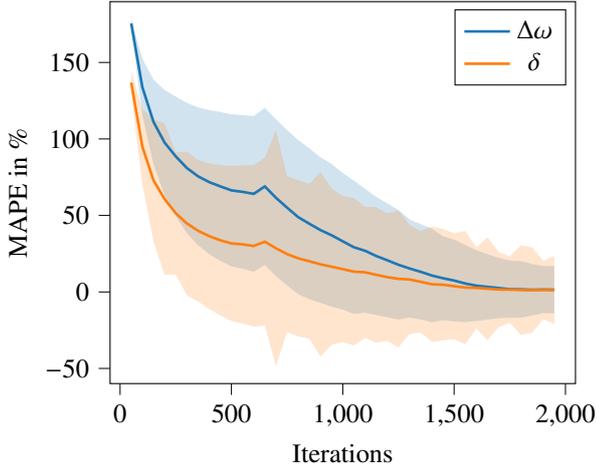
\begin{figure}
     \centering
\begin{tikzpicture}

\definecolor{darkgray176}{RGB}{176,176,176}
\definecolor{darkorange25512714}{RGB}{255,127,14}
\definecolor{steelblue31119180}{RGB}{31,119,180}

\begin{axis}[
width=.42\textwidth,
tick align=outside,
tick pos=left,
x grid style={darkgray176},
xlabel={Iterations},
xmin=-45, xmax=2045,
xtick style={color=black},
y grid style={darkgray176},
ylabel={MAPE in \(\displaystyle \%\)},
ymin=-59.8921853163442, ymax=189.753032531639,
ytick style={color=black}
]
\path [fill=steelblue31119180, fill opacity=0.2]
(axis cs:50,178.405522629458)
--(axis cs:50,172.553914533844)
--(axis cs:100,115.154377092887)
--(axis cs:150,83.4612177469218)
--(axis cs:200,63.1236345659918)
--(axis cs:250,49.4983249724718)
--(axis cs:300,38.347269447635)
--(axis cs:350,30.5339222578052)
--(axis cs:400,24.7716195581603)
--(axis cs:450,20.3221163534492)
--(axis cs:500,16.6714511771911)
--(axis cs:550,15.2145805212636)
--(axis cs:600,13.1927985946853)
--(axis cs:650,17.6212760238235)
--(axis cs:700,10.2501756663833)
--(axis cs:750,4.2910292306989)
--(axis cs:800,-1.55288255756911)
--(axis cs:850,-4.7127789575463)
--(axis cs:900,-7.1312418684465)
--(axis cs:950,-9.62032753933939)
--(axis cs:1000,-11.5417059474659)
--(axis cs:1050,-14.3647364824079)
--(axis cs:1100,-13.6168564900933)
--(axis cs:1150,-15.1591263624471)
--(axis cs:1200,-16.5322566845248)
--(axis cs:1250,-18.0554352027418)
--(axis cs:1300,-16.7356928105244)
--(axis cs:1350,-18.0193350524879)
--(axis cs:1400,-19.862929237214)
--(axis cs:1450,-18.4719710861573)
--(axis cs:1500,-19.3139769650073)
--(axis cs:1550,-19.72662009474)
--(axis cs:1600,-19.0213936783789)
--(axis cs:1650,-18.113827600465)
--(axis cs:1700,-17.1737972372916)
--(axis cs:1750,-16.8946742954912)
--(axis cs:1800,-16.7825040237086)
--(axis cs:1850,-15.4810534162906)
--(axis cs:1900,-13.8490611424839)
--(axis cs:1950,-13.9646448641024)
--(axis cs:1950,16.9454695539447)
--(axis cs:1950,16.9454695539447)
--(axis cs:1900,16.972605254019)
--(axis cs:1850,18.3197454119442)
--(axis cs:1800,20.1368658906213)
--(axis cs:1750,20.4857689626169)
--(axis cs:1700,22.3107140775448)
--(axis cs:1650,24.7598200912527)
--(axis cs:1600,27.2605489117012)
--(axis cs:1550,30.724637402162)
--(axis cs:1500,34.1961017865315)
--(axis cs:1450,36.3939010089643)
--(axis cs:1400,41.3265784441282)
--(axis cs:1350,44.3156700689668)
--(axis cs:1300,47.3786929597999)
--(axis cs:1250,53.561998039993)
--(axis cs:1200,57.8841970919322)
--(axis cs:1150,62.1750740331312)
--(axis cs:1100,67.4349081511932)
--(axis cs:1050,72.7396088204751)
--(axis cs:1000,77.732316076907)
--(axis cs:950,83.672253567731)
--(axis cs:900,87.9803303763124)
--(axis cs:850,93.8622537786736)
--(axis cs:800,99.3108705130112)
--(axis cs:750,105.86384475637)
--(axis cs:700,112.802158207653)
--(axis cs:650,120.332306762421)
--(axis cs:600,115.004706163115)
--(axis cs:550,115.608634359717)
--(axis cs:500,116.109798999346)
--(axis cs:450,117.659634821799)
--(axis cs:400,118.816891535395)
--(axis cs:350,120.645687911493)
--(axis cs:300,123.596229605859)
--(axis cs:250,127.697499646923)
--(axis cs:200,132.035920560736)
--(axis cs:150,138.792554446567)
--(axis cs:100,152.444032359498)
--(axis cs:50,178.405522629458)
--cycle;

\path [fill=darkorange25512714, fill opacity=0.2]
(axis cs:50,144.174367461197)
--(axis cs:50,129.530082581834)
--(axis cs:100,70.2595002329649)
--(axis cs:150,32.9887748943601)
--(axis cs:200,11.0633169478406)
--(axis cs:250,11.542306018392)
--(axis cs:300,-2.25900996457586)
--(axis cs:350,-6.2248405842098)
--(axis cs:400,-11.1912143911089)
--(axis cs:450,-15.3386636818764)
--(axis cs:500,-19.0388330990245)
--(axis cs:550,-20.6933557859208)
--(axis cs:600,-22.6813768968256)
--(axis cs:650,-22.0398106264773)
--(axis cs:700,-48.5446754141632)
--(axis cs:750,-26.2000201889225)
--(axis cs:800,-29.048251176811)
--(axis cs:850,-30.6515632435166)
--(axis cs:900,-42.2578594525133)
--(axis cs:950,-34.5822159011383)
--(axis cs:1000,-32.8504336622451)
--(axis cs:1050,-34.9733702575396)
--(axis cs:1100,-29.9531201660898)
--(axis cs:1150,-33.1424073757054)
--(axis cs:1200,-32.0723308196076)
--(axis cs:1250,-36.2148727563334)
--(axis cs:1300,-27.5243451838105)
--(axis cs:1350,-26.6258779104518)
--(axis cs:1400,-32.7457516147539)
--(axis cs:1450,-31.6854582569954)
--(axis cs:1500,-30.826178256998)
--(axis cs:1550,-34.2051128127269)
--(axis cs:1600,-23.8257709863826)
--(axis cs:1650,-31.4820859249435)
--(axis cs:1700,-22.993633927508)
--(axis cs:1750,-20.3007577834752)
--(axis cs:1800,-28.0108517047829)
--(axis cs:1850,-26.813943840382)
--(axis cs:1900,-17.7462419244767)
--(axis cs:1950,-20.9683586830373)
--(axis cs:1950,23.3675743441815)
--(axis cs:1950,23.3675743441815)
--(axis cs:1900,20.502381911935)
--(axis cs:1850,29.1632387750864)
--(axis cs:1800,30.5260564528669)
--(axis cs:1750,23.3393087569925)
--(axis cs:1700,26.1804240121723)
--(axis cs:1650,35.7212136612018)
--(axis cs:1600,29.0637812746961)
--(axis cs:1550,39.9363511466869)
--(axis cs:1500,38.3207247260303)
--(axis cs:1450,41.1364985461367)
--(axis cs:1400,42.8011294201562)
--(axis cs:1350,39.8479858268687)
--(axis cs:1300,43.8196479951772)
--(axis cs:1250,53.3302544906789)
--(axis cs:1200,51.38638359265)
--(axis cs:1150,55.5390811552021)
--(axis cs:1100,55.6496150422928)
--(axis cs:1050,61.4685890906461)
--(axis cs:1000,62.672614901161)
--(axis cs:950,67.572647420551)
--(axis cs:900,78.3303150111976)
--(axis cs:850,70.735442302687)
--(axis cs:800,72.9976212123479)
--(axis cs:750,75.6906034093443)
--(axis cs:700,105.63946359371)
--(axis cs:650,87.7321043502277)
--(axis cs:600,82.8048834303885)
--(axis cs:550,82.8602690858422)
--(axis cs:500,82.4575881789405)
--(axis cs:450,82.8812516657422)
--(axis cs:400,84.1325590617274)
--(axis cs:350,86.2022063404794)
--(axis cs:300,91.7612624193073)
--(axis cs:250,91.2749432968787)
--(axis cs:200,110.280421022963)
--(axis cs:150,112.743036822804)
--(axis cs:100,119.716599473328)
--(axis cs:50,144.174367461197)
--cycle;

\addplot [line width=1pt, steelblue31119180]
table {%
50 175.479718581651
100 133.799204726192
150 111.126886096744
200 97.5797775633638
250 88.5979123096973
300 80.9717495267471
350 75.5898050846492
400 71.7942555467776
450 68.9908755876242
500 66.3906250882683
550 65.4116074404904
600 64.0987523789002
650 68.9767913931223
700 61.5261669370182
750 55.0774369935344
800 48.8789939777211
850 44.5747374105636
900 40.4245442539329
950 37.0259630141958
1000 33.0953050647205
1050 29.1874361690336
1100 26.90902583055
1150 23.507973835342
1200 20.6759702037037
1250 17.7532814186256
1300 15.3215000746378
1350 13.1481675082395
1400 10.7318246034571
1450 8.96096496140349
1500 7.4410624107621
1550 5.49900865371098
1600 4.11957761666113
1650 3.32299624539387
1700 2.56845842012661
1750 1.79554733356283
1800 1.67718093345636
1850 1.41934599782678
1900 1.56177205576754
1950 1.49041234492114
};
\addlegendentry{$\Delta \omega$}
\addplot [line width=1pt, darkorange25512714]
table {%
50 136.852225021515
100 94.9880498531465
150 72.8659058585822
200 60.6718689854018
250 51.4086246576353
300 44.7511262273657
350 39.9886828781348
400 36.4706723353092
450 33.7712939919329
500 31.709377539958
550 31.0834566499607
600 30.0617532667814
650 32.8461468618752
700 28.5473940897736
750 24.7452916102109
800 21.9746850177684
850 20.0419395295852
900 18.0362277793421
950 16.4952157597063
1000 14.911090619458
1050 13.2476094165533
1100 12.8482474381015
1150 11.1983368897483
1200 9.65702638652121
1250 8.55769086717278
1300 8.14765140568334
1350 6.61105395820846
1400 5.02768890270113
1450 4.72552014457064
1500 3.74727323451613
1550 2.86561916698001
1600 2.61900514415676
1650 2.11956386812917
1700 1.59339504233212
1750 1.51927548675865
1800 1.25760237404198
1850 1.17464746735222
1900 1.3780699937291
1950 1.19960783057209
};
\addlegendentry{$\delta$}
\end{axis}

\end{tikzpicture}
     \caption{Influence of iterations on estimation error MAPE with $2\,\sigma$ ($2\,\sigma$ is represented by the colored areas, training and estimations performed on 118-bus scenario ”fast dynamics” )}
     \label{fig:MAPE_over_iterations}
 \end{figure}

\paragraph{Reduction of training iterations}
\cref{fig:MAPE_over_iterations} shows the evolution of MAPE over iterations for training without the pretraining. It takes around 1500 iterations to reach the final MAPE.
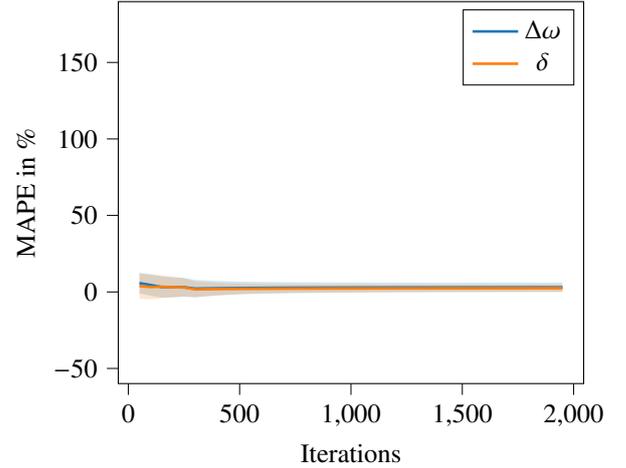
\begin{figure}
    \centering
\begin{tikzpicture}

\definecolor{darkgray176}{RGB}{176,176,176}
\definecolor{darkorange25512714}{RGB}{255,127,14}
\definecolor{steelblue31119180}{RGB}{31,119,180}

\begin{axis}[
width=.42\textwidth,
tick align=outside,
tick pos=left,
x grid style={darkgray176},
xlabel={Iterations},
xmin=-45, xmax=2045,
ymin=-59.8921853163442, ymax=189.753032531639,
xtick style={color=black},
y grid style={darkgray176},
ylabel={MAPE in \(\displaystyle \%\)},
ytick style={color=black}
]
\path [fill=steelblue31119180, fill opacity=0.2]
(axis cs:50,12.4118768020985)
--(axis cs:50,-0.851876802098476)
--(axis cs:100,-2.4846132974845)
--(axis cs:150,-3.73884296347968)
--(axis cs:200,-3.49305166461602)
--(axis cs:250,-3.02681989566645)
--(axis cs:300,-3.45076405299387)
--(axis cs:350,-2.85742423972107)
--(axis cs:400,-2.3566267090274)
--(axis cs:450,-1.96473569136522)
--(axis cs:500,-1.65487582263162)
--(axis cs:550,-1.40931849465535)
--(axis cs:600,-1.21103968184084)
--(axis cs:650,-1.05050058413854)
--(axis cs:700,-0.919599762230929)
--(axis cs:750,-0.811158944728213)
--(axis cs:800,-0.719219234094739)
--(axis cs:850,-0.639847019612084)
--(axis cs:900,-0.570960079583704)
--(axis cs:950,-0.510175771274777)
--(axis cs:1000,-0.456671036285787)
--(axis cs:1050,-0.408051154668251)
--(axis cs:1100,-0.364245158098619)
--(axis cs:1150,-0.324304740125563)
--(axis cs:1200,-0.288995827127886)
--(axis cs:1250,-0.256099213946428)
--(axis cs:1300,-0.226493084843911)
--(axis cs:1350,-0.19923147324991)
--(axis cs:1400,-0.173830603285936)
--(axis cs:1450,-0.150809625785915)
--(axis cs:1500,-0.129888546268481)
--(axis cs:1550,-0.110583917197136)
--(axis cs:1600,-0.0929249694362397)
--(axis cs:1650,-0.0767379829878561)
--(axis cs:1700,-0.0623304157466991)
--(axis cs:1750,-0.0490975443427946)
--(axis cs:1800,-0.0368899185567404)
--(axis cs:1850,-0.0257728481023078)
--(axis cs:1900,-0.0156710763407242)
--(axis cs:1950,-0.00652867304677462)
--(axis cs:1950,6.03305817211163)
--(axis cs:1950,6.03305817211163)
--(axis cs:1900,6.0303424316741)
--(axis cs:1850,6.02836543920982)
--(axis cs:1800,6.02728406597252)
--(axis cs:1750,6.02721691359789)
--(axis cs:1700,6.02793291972382)
--(axis cs:1650,6.02904079076594)
--(axis cs:1600,6.03179377617992)
--(axis cs:1550,6.03581607138641)
--(axis cs:1500,6.04121473606383)
--(axis cs:1450,6.0481716814572)
--(axis cs:1400,6.05666005272618)
--(axis cs:1350,6.06701892343265)
--(axis cs:1300,6.07854686069938)
--(axis cs:1250,6.09170581225789)
--(axis cs:1200,6.10668398848891)
--(axis cs:1150,6.12256347967959)
--(axis cs:1100,6.1411601013229)
--(axis cs:1050,6.16154548005168)
--(axis cs:1000,6.18376152785478)
--(axis cs:950,6.20768868588503)
--(axis cs:900,6.23413303702729)
--(axis cs:850,6.26344627577338)
--(axis cs:800,6.29785605134157)
--(axis cs:750,6.33795416710817)
--(axis cs:700,6.38649533597269)
--(axis cs:650,6.4462814087641)
--(axis cs:600,6.52270928731659)
--(axis cs:550,6.62307239429513)
--(axis cs:500,6.75635240039239)
--(axis cs:450,6.93246671086141)
--(axis cs:400,7.16824441970725)
--(axis cs:350,7.47945327505434)
--(axis cs:300,7.83782405299387)
--(axis cs:250,9.24881989566645)
--(axis cs:200,9.71505166461602)
--(axis cs:150,10.1788429634797)
--(axis cs:100,11.5956132974845)
--(axis cs:50,12.4118768020985)
--cycle;

\path [fill=darkorange25512714, fill opacity=0.2]
(axis cs:50,12.1139223774585)
--(axis cs:50,-4.41392237745849)
--(axis cs:100,-4.8244894811562)
--(axis cs:150,-3.98818949739386)
--(axis cs:200,-3.59907969637379)
--(axis cs:250,-3.05963413683138)
--(axis cs:300,-3.54959582287673)
--(axis cs:350,-2.78599687572507)
--(axis cs:400,-2.16416184062456)
--(axis cs:450,-1.69519906599288)
--(axis cs:500,-1.33992617329479)
--(axis cs:550,-1.0695449378294)
--(axis cs:600,-0.863378770901083)
--(axis cs:650,-0.703508577105495)
--(axis cs:700,-0.578893816770599)
--(axis cs:750,-0.480491587391828)
--(axis cs:800,-0.400680941792304)
--(axis cs:850,-0.336361201026442)
--(axis cs:900,-0.282662767802789)
--(axis cs:950,-0.237802974418439)
--(axis cs:1000,-0.199614005556267)
--(axis cs:1050,-0.166675353129652)
--(axis cs:1100,-0.138048253417579)
--(axis cs:1150,-0.11313379779667)
--(axis cs:1200,-0.0915294683057706)
--(axis cs:1250,-0.0722698678244496)
--(axis cs:1300,-0.0548990764541513)
--(axis cs:1350,-0.0397724266161079)
--(axis cs:1400,-0.0261650218301912)
--(axis cs:1450,-0.0139799275900154)
--(axis cs:1500,-0.00330716066241843)
--(axis cs:1550,0.00637736075224549)
--(axis cs:1600,0.0149972049669325)
--(axis cs:1650,0.0228820648425185)
--(axis cs:1700,0.030329509211942)
--(axis cs:1750,0.0365095845032357)
--(axis cs:1800,0.0423570701470388)
--(axis cs:1850,0.0478232318625285)
--(axis cs:1900,0.0523582253511816)
--(axis cs:1950,0.0564970461207528)
--(axis cs:1950,4.81960075480323)
--(axis cs:1950,4.81960075480323)
--(axis cs:1900,4.81107613304104)
--(axis cs:1850,4.80269298942778)
--(axis cs:1800,4.79503745110564)
--(axis cs:1750,4.78781402509247)
--(axis cs:1700,4.78065800837018)
--(axis cs:1650,4.7738326048675)
--(axis cs:1600,4.76727865853266)
--(axis cs:1550,4.7615109972163)
--(axis cs:1500,4.75654009995614)
--(axis cs:1450,4.75232109530629)
--(axis cs:1400,4.7486532049925)
--(axis cs:1350,4.74596558028899)
--(axis cs:1300,4.74414390138706)
--(axis cs:1250,4.74401260158297)
--(axis cs:1200,4.74465170584624)
--(axis cs:1150,4.74660383160261)
--(axis cs:1100,4.75012074236428)
--(axis cs:1050,4.755619964061)
--(axis cs:1000,4.76294866159776)
--(axis cs:950,4.7729160321733)
--(axis cs:900,4.78611642003415)
--(axis cs:850,4.80420250563062)
--(axis cs:800,4.82804814622862)
--(axis cs:750,4.86035108425683)
--(axis cs:700,4.90519526337307)
--(axis cs:650,4.96734433586663)
--(axis cs:600,5.05415968507985)
--(axis cs:550,5.17677454008185)
--(axis cs:500,5.34906943912636)
--(axis cs:450,5.58953261643423)
--(axis cs:400,5.92214295821144)
--(axis cs:350,6.37915386273821)
--(axis cs:300,6.9343465373187)
--(axis cs:250,8.85943413683138)
--(axis cs:200,9.82107969637379)
--(axis cs:150,10.8981894973939)
--(axis cs:100,11.2684894811562)
--(axis cs:50,12.1139223774585)
--cycle;

\addplot [line width=1pt, steelblue31119180]
table {%
50 5.78
100 4.5555
150 3.22
200 3.111
250 3.111
300 2.19353
350 2.31101451766664
400 2.40580885533992
450 2.48386550974809
500 2.55073828888038
550 2.60687694981989
600 2.65583480273788
650 2.69789041231278
700 2.73344778687088
750 2.76339761118998
800 2.78931840862342
850 2.81179962808065
900 2.83158647872179
950 2.84875645730513
1000 2.8635452457845
1050 2.87674716269171
1100 2.88845747161214
1150 2.89912936977701
1200 2.90884408068051
1250 2.91780329915573
1300 2.92602688792774
1350 2.93389372509137
1400 2.94141472472012
1450 2.94868102783564
1500 2.95566309489767
1550 2.96261607709464
1600 2.96943440337184
1650 2.97615140388904
1700 2.98280125198856
1750 2.98905968462755
1800 2.99519707370789
1850 3.00129629555376
1900 3.00733567766669
1950 3.01326474953243
};
\addlegendentry{$\Delta \omega$}
\addplot [line width=1pt, darkorange25512714]
table {%
50 3.85
100 3.222
150 3.455
200 3.111
250 2.8999
300 1.69237535722099
350 1.79657849350657
400 1.87899055879344
450 1.94716677522068
500 2.00457163291578
550 2.05361480112623
600 2.09539045708938
650 2.13191787938057
700 2.16315072330123
750 2.1899297484325
800 2.21368360221816
850 2.23392065230209
900 2.25172682611568
950 2.26755652887743
1000 2.28166732802075
1050 2.29447230546568
1100 2.30603624447335
1150 2.31673501690297
1200 2.32656111877023
1250 2.33587136687926
1300 2.34462241246645
1350 2.35309657683644
1400 2.36124409158115
1450 2.36917058385814
1500 2.37661646964686
1550 2.38394417898427
1600 2.3911379317498
1650 2.39835733485501
1700 2.40549375879106
1750 2.41216180479785
1800 2.41869726062634
1850 2.42525811064515
1900 2.43171717919611
1950 2.43804890046199
};
\addlegendentry{$\delta$}

\end{axis}

\end{tikzpicture}
    \caption{Transfer learning MAPE over iterations with $2\sigma$ ($2\,\sigma$ is represented by the colored areas, pretrained on SMIB "slow dynamics", training and estimations performed on 118-bus scenario "fast dynamics")}
    \label{fig:MAPE_iterations_transfer}
\end{figure}
In \cref{fig:MAPE_iterations_transfer} $\text{MAPE}_{\Delta \omega}$ and $\text{MAPE}_{\delta}$ are shown over the number of iterations for training on the 118-bus system after pretraining on the \ac{smib}. The final $\text{MAPE}_{\Delta \omega}$ is reached after 500 iterations, which is a reduction in training iterations of $75\,\%$ compared to the estimations performed previously. The final $\text{MAPE}_{\delta}$ is reached after 550 iterations, which still reduces the training iterations by $72.5\,\%$. In conclusion, the results in \cref{fig:MAPE_over_iterations} and \cref{fig:MAPE_iterations_transfer} indicate that the \ac{bpinn} is capable of transferring previously learned knowledge to the new space. 

\begin{figure}
    \centering
\begin{tikzpicture}

\definecolor{cadetblue96170144}{RGB}{96,170,144}
\definecolor{darkgray176}{RGB}{176,176,176}
\definecolor{darkseagreen164204144}{RGB}{164,204,144}
\definecolor{lightgray204}{RGB}{204,204,204}
\definecolor{midnightblue4448113}{RGB}{44,48,113}
\definecolor{steelblue51132141}{RGB}{51,132,141}
\definecolor{teal2993134}{RGB}{29,93,134}

\begin{axis}[
legend cell align={left},
legend style={
  fill opacity=0.8,
  draw opacity=1,
  text opacity=1,
  at={(1,0.5)},
  anchor=west,
  draw=lightgray204
},
width=.4\textwidth,
tick align=outside,
tick pos=left,
x grid style={darkgray176},
xlabel={\(\displaystyle N_z\)},
xticklabels={, 50,,,, , 10,,,, ,5},
xmin=-0.1, xmax=2.1,
y grid style={darkgray176},
ylabel={MAPE in \(\displaystyle \%\)},
ymin=-5, ymax=120,
name=first axis,
anchor=south,
]
\addlegendimage{empty legend}\addlegendentry{$x$}
\addlegendimage{mark=*, only marks}\addlegendentry{$\Delta \omega$}
\addlegendimage{draw=black, mark=x, only marks}\addlegendentry{$\delta$}
\addlegendimage{empty legend}\addlegendentry{$N_c$}
\addplot [ draw=white, fill=darkseagreen164204144, mark=*, only marks]
coordinates{(0.0, 2.5466266222591987) (1.0, 21.352100902301878)(2.0, 80.41582431744708)
};

\addlegendentry{$0 \cdot N_z$}
\addplot [draw=white, fill=cadetblue96170144, mark=*, only marks]
coordinates{(0.0, 2.763397611189979) (1.0, 2.1526308286815388)(2.0, 21.352100902301878)};

\addlegendentry{$1\cdot N_z$}
\addplot [draw=white, fill=steelblue51132141, mark=*, only marks]
coordinates{(0.0, 2.73092477431835) (1.0, 2.3640194027953294)(2.0, 2.530806835746906)};

\addlegendentry{$2\cdot N_z$}
\addplot [draw=white, fill=teal2993134, mark=*, only marks]
coordinates{(0.0, 2.842142017319502) (1.0, 2.402066114084384) (2.0,2.1526308286815388 )
};

\addlegendentry{$3\cdot N_z$}
\addplot [draw=white, fill=midnightblue4448113, mark=*, only marks]
coordinates{
(0.0, 2.7175413006077145) (1.0, 2.5466266222591987)  (2.0,2.26086373620591 )};

\addlegendentry{$4\cdot N_z$}

\addplot [ draw=darkseagreen164204144, fill=darkseagreen164204144, mark=x, only marks]
coordinates{(0.0, 1.9759562435173617) (1.0, 9.776223440685552)(2.0, 95.59477532881662)
};

\addplot [draw=cadetblue96170144, fill=cadetblue96170144, mark=x, only marks]
coordinates{(0.0, 2.1899297484324998) (1.0, 1.6967152332750322)(2.0, 9.776223440685552)};

\addplot [draw=steelblue51132141, fill=steelblue51132141, mark=x, only marks]
coordinates{(0.0, 2.1734970860366456) (1.0, 1.7979014413858967)(2.0, 2.637217672115641)};

\addplot [draw=teal2993134, fill=teal2993134, mark=x, only marks]
coordinates{(0.0, 2.307249597846985) (1.0, 1.8425902718528813) (2.0,1.6967152332750322)};

\addplot [draw=midnightblue4448113, fill=midnightblue4448113, mark=x, only marks]
coordinates{(0.0, 2.178684663397885) (1.0, 1.9759562435173617)  (2.0,1.7190067581260189	 )};

\end{axis}

\end{tikzpicture}
    \caption{Transfer learning MAPE over step size (of IEEE 118-bus data) and collocation points (trained on SMIB "slow dynamics", estimations performed on 118-bus scenario "fast dynamics")}
    \label{fig:MAPE_step_size_coll_transfer_learning}
\end{figure}
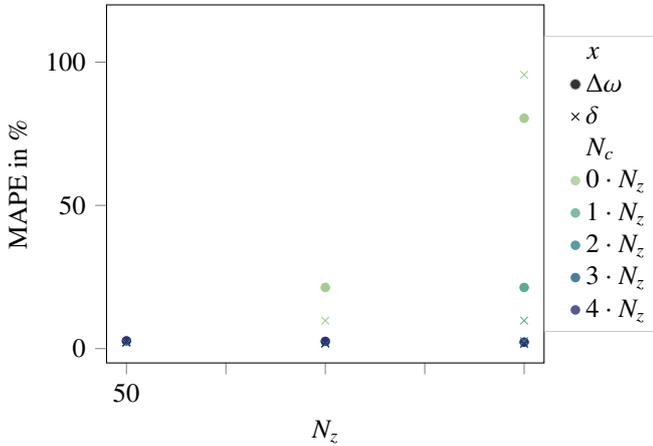

\paragraph{Reduction of data samples}
This paragraph explores the transfer learning performance for sparse data. The aim is to achieve a performance comparable to full training in the target domain, i.e. \cref{fig:mape_step_size_coll}, with pretraining and less data. Similarly to the previous analysis, we also enrich the data with collocation points. \cref{fig:MAPE_step_size_coll_transfer_learning} demonstrates that a decrease in samples $N_z$ results in an increased \ac{mape} for both quantities similar to \cref{fig:mape_step_size_coll}. 
Augmenting the data with $N_c=1\cdot N_z$ collocation points already leads to a significant improvement in the estimation accuracy for $N_z=10$ and $N_z=5$. However, compared to \cref{fig:mape_step_size_coll}, it can be seen that pretraining and transfer learning reduces the error, so $N_z=10$ samples augmented with 10 collocation points already lead to the target error. Similarly, the amount of data required to reach the target error can be reduced for $N_z=5$. After that, there are no improvements in the estimation error. 
The \ac{bpinn} is able to achieve the target error with $N_z=5$ and $N_c=2\cdot N_z$, which is significantly less data than in \cref{fig:mape_step_size_coll}. We can conclude from these results that the \ac{bpinn} benefits from pretraining and transfer learning. This allows us to significantly reduce the training iterations and also slightly the number of collocation points in case of sparse data.

\subsection{Discussion}
\paragraph{\ac{bpinn} vs. \ac{pinn} vs. \ac{sindy}}
The results indicate that the \ac{bpinn} and \ac{pinn} cannot achieve similar results to \ac{sindy} for the \ac{smib} system. In that case, the power system is fully represented by the regression formulation utilized \cref{eq:regression_problem}. This advances \ac{sindy} due to its point-wise fitting approach. However, the \ac{bpinn} and \ac{pinn} achieve errors in the range of a few percents, which is acceptable. This reverses in the presence of \acp{ibr}, i.e. 3-bus, CIGRE 14-bus and IEEE 118-bus systems. The regression formulation \cref{eq:regression_problem} no longer represents the model behavior to the full extent. In this situation, it turns out to be beneficial that the \ac{bpinn} fits a distribution and obtains the system parameters $\boldsymbol{\lambda}$ based on its learned surrogate model. The \ac{pinn} also learns a surrogate model, but in contrast seeks to find a single best estimate for the system parameters $\boldsymbol{\lambda}$, which appears to be detrimental in this case study. 

We also aimed to compare the \ac{bpinn} with the unscented Kalman filter throughout our investigations. However, this approach requires prior assumptions of process uncertainty and measurement noise matrices. We found these to be significantly different for various system dynamics. In a fast-changing system, such as the power system with high shares of \acp{ibr}, this would not be appropriate.  

\paragraph{$\text{MAPE}_{\Delta \omega}$ vs. $\text{MAPE}_{\delta}$}
The results show that the \ac{bpinn}s $\text{MAPE}_\delta$ error is often smaller than the $\text{MAPE}_{\Delta \omega}$. This stems from the shape of the individual trajectories. The $\Delta\omega$ trajectories show a more complex behavior, deviating around zero, whereas the $\delta$ trajectories move from the initial angle to the new angle. The absolute values of the individual quantities $\boldsymbol{x}$ are not influential, as they are normalized before the estimation process.

\paragraph{Uncertainty quantification}
The \cref{tbl:Noise_results} reveals that the posterior standard deviation differs significantly for all systems in the same scenario. We expect a wider distribution in the presence of epistemic uncertainty. This expectation is based on the formulation of the quantified uncertainty \cref{eq:uncertainty_quant}, which depends on the epistemic uncertainty. However, in our simulation, another factor comes into play that influences the width of the posterior distribution. We previously found that slower dynamics potentially lead to increased estimation errors and consequently wider posterior standard deviation. Consequently, the uncertainty, represented by the posterior standard deviation, depends on the dynamics of the system, and the share of \acp{ibr}. Both factors also influence each other, which complicates the interpretation of the posterior standard deviation as a confidence measure. The distinction between different sources of uncertainty can only be made in theory. In practice, the \ac{bpinn} would give one confidence measure and the interpretation requires significant experience.

\paragraph{Runtime}
All experiments were performed on an Intel i7 11700 CPU. The average training time of the \ac{bpinn} over all scenarios and systems was $18\,\text{s}$ for 2000 iterations with a power system simulation step size of $T_s=0.05\,\text{s}$. Similar training times were achieved for the \ac{pinn} with an average runtime of $17.8\,\text{s}$. \ac{sindy} training took $0.0027\,\text{s}$ on average for one estimate. It should be noted here that the \ac{bpinn} presents its estimate as a distribution, which can be seen as computationally equivalent to performing multiple single estimates at the same time. Nevertheless, there is potential to improve the \ac{bpinn} estimation speed, for example with GPU utilization.

\section{Conclusion}
\label{sec:conclusion}
In this paper, we explored the \acf{bpinn} for system identification under model uncertainties resulting from \acfp{ibr}. We evaluated the performance in four different grids: the \ac{smib}, a 3-bus system, CIGRE 14-bus and IEEE 118-bus system equipped with multiple \acp{ibr}. The \ac{bpinn} achieved lower estimation errors compared to the widely popular system identification method \ac{sindy} by a factor of $10$ up to $90$ in presence of \acp{ibr} and factor $2$ to $3$ compared to the \ac{pinn}. In addition, we found that transfer learning is beneficial in \ac{bpinn} training to reduce the number of iterations and the amount of required data. Pretraining on a \ac{smib} system reduces the training time by up to $75\,\%$ for estimation on the 118-bus system. The amount of required collocation points can also be reduced by pretraining and transfer learning.

\bibliographystyle{elsarticle-num} 
\bibliography{library.bib}

\begin{thebibliography}{10}
\expandafter\ifx\csname url\endcsname\relax
  \def\url#1{\texttt{#1}}\fi
\expandafter\ifx\csname urlprefix\endcsname\relax\def\urlprefix{URL }\fi
\expandafter\ifx\csname href\endcsname\relax
  \def\href#1#2{#2} \def\path#1{#1}\fi

\bibitem{zhao2019}
J.~Zhao, A.~Gómez-Expósito, M.~Netto, L.~Mili, A.~Abur, V.~Terzija, I.~Kamwa,
  B.~Pal, A.~K. Singh, J.~Qi, Z.~Huang, A.~P.~S. Meliopoulos, Power system
  dynamic state estimation: Motivations, definitions, methodologies, and future
  work, IEEE Transactions on Power Systems 34~(4) (2019) 3188--3198.
\newblock \href {https://doi.org/10.1109/TPWRS.2019.2894769}
  {\path{doi:10.1109/TPWRS.2019.2894769}}.

\bibitem{Susuki2018}
Y.~Susuki, R.~Hamasaki, A.~Ishigame, Estimation of power system inertia using
  nonlinear koopman modes, in: 2018 IEEE Power $\&$ Energy Society General
  Meeting (PESGM), 2018, pp. 1--5.
\newblock \href {https://doi.org/10.1109/PESGM.2018.8586007}
  {\path{doi:10.1109/PESGM.2018.8586007}}.

\bibitem{Brunton2016}
S.~L. Brunton, J.~L. Procter, J.~N. Kutz, Discovering governing equations from
  data by sparse identification of nonlinear dynamical systems, Applied
  Mathematics 113~(15) (2016) 3932--3937.
\newblock \href {https://doi.org/10.1073/pnas.1517384113}
  {\path{doi:10.1073/pnas.1517384113}}.

\bibitem{Stiasny2020}
J.~Stiasny, G.~S. Misyris, S.~Chatzivasileiadis, Physics-informed neural
  networks for non-linear system identification for power system dynamics, in:
  2021 IEEE Madrid PowerTech, 2021, pp. 1--6.
\newblock \href {https://doi.org/10.1109/PowerTech46648.2021.9495063}
  {\path{doi:10.1109/PowerTech46648.2021.9495063}}.

\bibitem{Petra2017}
N.~Petra, C.~G. Petra, Z.~Zhang, E.~M. Constantinescu, M.~Anitescu, A bayesian
  approach for parameter estimation with uncertainty for dynamic power systems,
  IEEE Transactions on Power Systems 32~(4) (2017) 2735--2743.
\newblock \href {https://doi.org/10.1109/TPWRS.2016.2625277}
  {\path{doi:10.1109/TPWRS.2016.2625277}}.

\bibitem{YANG2021}
L.~Yang, X.~Meng, G.~E. Karniadakis, B-pinns: Bayesian physics-informed neural
  networks for forward and inverse pde problems with noisy data, Journal of
  Computational Physics 425 (2021) 109913.
\newblock \href {https://doi.org/https://doi.org/10.1016/j.jcp.2020.109913}
  {\path{doi:https://doi.org/10.1016/j.jcp.2020.109913}}.

\bibitem{Stock2023}
S.~Stock, J.~Stiasny, D.~Babazadeh, C.~Becker, S.~Chatzivasileiadis, Bayesian
  physics-informed neural networks for robust system identification of power
  systems, in: 2023 IEEE Belgrade PowerTech, 2023, pp. 1--6.
\newblock \href {https://doi.org/10.1109/PowerTech55446.2023.10202692}
  {\path{doi:10.1109/PowerTech55446.2023.10202692}}.

\bibitem{RAISSI2019}
M.~Raissi, P.~Perdikaris, G.~Karniadakis, Physics-informed neural networks: A
  deep learning framework for solving forward and inverse problems involving
  nonlinear partial differential equations, Journal of Computational Physics
  378 (2019) 686--707.
\newblock \href {https://doi.org/https://doi.org/10.1016/j.jcp.2018.10.045}
  {\path{doi:https://doi.org/10.1016/j.jcp.2018.10.045}}.

\bibitem{Kendall2017}
A.~Kendall, Y.~Gal, What uncertainties do we need in bayesian deep learning for
  computer vision?, in: I.~Guyon, U.~V. Luxburg, S.~Bengio, H.~Wallach,
  R.~Fergus, S.~Vishwanathan, R.~Garnett (Eds.), Advances in Neural Information
  Processing Systems, Vol.~30, Curran Associates, Inc., 2017.

\bibitem{Kononenko1989}
I.~Kononenko, Bayesian neural networks, Biological Cybernetics 61~(5) (1989)
  361--370.
\newblock \href {https://doi.org/10.1007/BF00200801}
  {\path{doi:10.1007/BF00200801}}.

\bibitem{graf2022}
O.~Graf, P.~Flores, P.~Protopapas, K.~Pichara, Error-aware b-pinns: Improving
  uncertainty quantification in bayesian physics-informed neural networks
  (2022).
\newblock \href {http://arxiv.org/abs/2212.06965} {\path{arXiv:2212.06965}}.

\bibitem{LIU2016}
Q.~Liu, D.~Wang, Stein variational gradient descent: A general purpose bayesian
  inference algorithm, in: D.~Lee, M.~Sugiyama, U.~Luxburg, I.~Guyon,
  R.~Garnett (Eds.), Advances in Neural Information Processing Systems,
  Vol.~29, Curran Associates, Inc., 2016.

\bibitem{Lemoine2019}
N.~P. Lemoine, Moving beyond noninformative priors: why and how to choose
  weakly informative priors in bayesian analyses, Oikos 128~(7) (2019)
  912--928.
\newblock \href {https://doi.org/https://doi.org/10.1111/oik.05985}
  {\path{doi:https://doi.org/10.1111/oik.05985}}.

\bibitem{Bergen1981}
A.~Bergen, D.~Hill, A structure preserving model for power system stability
  analysis, IEEE Transactions on Power Apparatus and Systems PAS-100~(1) (1981)
  25--35.
\newblock \href {https://doi.org/10.1109/TPAS.1981.316883}
  {\path{doi:10.1109/TPAS.1981.316883}}.

\bibitem{Kaptanoglu2022}
A.~Kaptanoglu, B.~de~Silva, U.~Fasel, K.~Kaheman, A.~Goldschmidt, J.~Callaham,
  C.~Delahunt, Z.~Nicolaou, K.~Champion, J.-C. Loiseau, J.~Kutz, S.~Brunton,
  Pysindy: A comprehensive python package for robust sparse system
  identification, Journal of Open Source Software 7~(69) (2022).

\bibitem{Zhong2011}
Q.-C. Zhong, G.~Weiss, Synchronverters: Inverters that mimic synchronous
  generators, IEEE Transactions on Industrial Electronics 58~(4) (2011)
  1259--1267.
\newblock \href {https://doi.org/10.1109/TIE.2010.2048839}
  {\path{doi:10.1109/TIE.2010.2048839}}.

\bibitem{Rosso2017}
R.~Rosso, J.~Cassoli, S.~Engelken, G.~Buticchi, M.~Liserre, Analysis and design
  of lcl filter based synchronverter, in: 2017 IEEE Energy Conversion Congress
  and Exposition (ECCE), 2017, pp. 5587--5594.
\newblock \href {https://doi.org/10.1109/ECCE.2017.8096930}
  {\path{doi:10.1109/ECCE.2017.8096930}}.

\bibitem{PAN2010}
S.~J. Pan, Q.~Yang, A survey on transfer learning, IEEE Transactions on
  Knowledge and Data Engineering 22~(10) (2010) 1345--1359.
\newblock \href {https://doi.org/10.1109/TKDE.2009.191}
  {\path{doi:10.1109/TKDE.2009.191}}.

\end{thebibliography}

\end{document}